\documentclass[9pt,twocolumn,twoside,a4paper]{extarticle}

\RequirePackage{xifthen}
\newboolean{arxiv}
\setboolean{arxiv}{true}

\ifthenelse{\boolean{arxiv}}{
 \RequirePackage[utf8]{inputenc}
 \RequirePackage{amsmath,amsfonts,amssymb}
 \RequirePackage{mathpazo} % Widely available alternative to Garamond
 \RequirePackage[scaled]{helvet}
 \RequirePackage[T1]{fontenc}
 \RequirePackage{url}
 \RequirePackage{xcolor}
 \definecolor{darkgreen}{RGB}{0,96,0}
 \definecolor{darkblue}{RGB}{0,0,96}
 \RequirePackage[colorlinks=true, linkcolor=darkblue, citecolor=darkgreen, filecolor=blue, urlcolor=black]{hyperref}
 
 \newcommand{\ociscodes}{}
 \usepackage{algorithm}
 
 \RequirePackage[numbers,sort&compress,super]{natbib}
 \setlength{\bibsep}{0.0pt}
 \bibliographystyle{unsrt}
 
 \usepackage{authblk}
 \setlength{\affilsep}{1em}

 \definecolor{color0}{RGB}{0,0,0} % Base
 \definecolor{color1}{RGB}{59,90,198} % author email, doi
 \definecolor{color2}{RGB}{0,96,0} % Header, authors, table lines

 \RequirePackage[a4paper,left=48pt,%
 right=42pt,%
 top=46pt,%
 bottom=60pt,%
 headheight=15pt,%
 headsep=10pt,%
 twoside]{geometry}%
 
 \RequirePackage{titlesec}
 \titleformat{\section}{\LARGE\sffamily\bfseries}{\thesection}{1em}{}
 \titlespacing{\section}{0em}{1em}{0em}
 \titleformat{\subsection}{\Large\sffamily\bfseries}{\thesubsection}{1em}{}
 \titlespacing{\subsection}{0em}{1em}{0em}
 \titleformat{\subsubsection}{\large\sffamily\bfseries}{\thesubsubsection}{1em}{}
 \titlespacing{\subsubsection}{0em}{1em}{0em}

}{
  \setboolean{shortarticle}{false}
  \setboolean{minireview}{false}
}

\RequirePackage{jabbrv}

\usepackage{todonotes}
\usepackage{upgreek}
\usepackage{xfrac}
\usepackage{algorithmicx}
\usepackage[english]{babel}
\usepackage{steinmetz} % for complex argument with \phase{}
\usepackage{algpseudocode}
\usepackage{algorithmicx}

\usepackage{xr} % To link back to the primary document

\newcommand{\mum}{\,\mathrm{\upmu m}} % non-italic mu only possible with Times roman font

\newcommand{\ri}{\mathrm{i}}
\newcommand{\rr}{\mathrm{r}}
\newcommand{\re}{\mathrm{e}}

\newcommand{\argmin}[1]{\underset{#1}{\operatorname{argmin}}}

\newcommand{\bx}{\boldsymbol{x}}

\newcommand{\bk}{\boldsymbol{k}}
\newcommand{\bn}{\boldsymbol{n}}
\newcommand{\bD}{\boldsymbol{\Delta}}
\newcommand{\bG}{\overset{\text{\tiny$\leftrightarrow$}}{\boldsymbol{G}}}
\renewcommand{\Re}{\mathrm{Re}}
\renewcommand{\Im}{\mathrm{Im}}

\newcommand{\iter}{p}
\newcommand{\maxeig}{{\max}}
\newcommand{\iterend}{N}
\newcommand{\nbpoints}{P}
\newcommand{\Op}{\boldsymbol{\mathcal{O}}}
\newcommand{\CurlOp}{\boldsymbol{\mathcal{D}}}
\newcommand{\sigmaCurl}{\sigma_\mathcal{D}}

\newcommand{\mdot}{} % We should reserver the dot product for multiplying vectors

\DeclareSymbolFont{bbold}{U}{bbold}{m}{n}
\DeclareSymbolFontAlphabet{\mathbbold}{bbold}
\newcommand{\one}{\mathbbold{1}}

\algnewcommand{\LineComment}[1]{\item[]\hfill\textcolor[rgb]{0,0.4,0}{ \(\triangleright\) \textit{#1}}}

\graphicspath{{figures/}}

\newboolean{supplement}
\setboolean{supplement}{false}

\setboolean{supplement}{true}
\begin{document}
\newcommand{\manuscripttitle}{Calculating coherent light-wave propagation\\ in large heterogeneous media}

\ifthenelse{\boolean{supplement}}{
  \ifthenelse{\boolean{arxiv}}{
  }{
    \setboolean{displaycopyright}{false}
  }
  
%  \title{\manuscripttitle: supplementary material}
%  \title{Calculating coherent light-wave propagation in\\ large heterogeneous media: supplementary material}
  \title{Supplementary Information:\\Calculating coherent light-wave propagation\\ in large heterogeneous media.}
  
  \renewcommand{\thepage}{S\arabic{page}}  
  \renewcommand{\thesection}{S\arabic{section}}
  \renewcommand{\thesubsection}{S\arabic{section}\Alph{subsection}}
  \renewcommand{\thesubsubsection}{S\arabic{section}\Alph{subsection}\Roman{subsubsection}}
  \renewcommand{\thetable}{S\arabic{table}}   
  \renewcommand{\thefigure}{S\arabic{figure}}
  \renewcommand{\thealgorithm}{S\arabic{algorithm}}
  \renewcommand{\theequation}{S\arabic{equation}} 
  
  \renewcommand\refname{Supplementary References}
}{
  \title{\manuscripttitle}
}

\author[1,2,*]{T.~Vettenburg}
\author[1]{S.A.R.~Horsley}
\author[1]{J.~Bertolotti}

\affil[1]{Department of Physics and Astronomy, University of Exeter, Exeter, EX4 4QL, United Kingdom}
\affil[2]{School of Science and Engineering, University of Dundee, Nethergate, Dundee, DD1 4HN, United Kingdom}
\affil[*]{Corresponding author: \href{mailto:t.vettenburg@dundee.ac.uk}{t.vettenburg@dundee.ac.uk}}

\ifthenelse{\boolean{arxiv}}{
  \date{December 25, 2018}
}{
  % To be edited by editor
  \dates{Compiled \today}
  
  %\ociscodes{(050.1755) Computational electromagnetic methods; (080.1753) Computation methods; (080.2720) Mathematical methods (general); (160.1190) Anisotropic optical materials; (260.1440) Birefringence; (160.1585) Chiral media; (160.3918) Metamaterials; (080.2710) Inhomogeneous optical media; (110.7050) Turbid media; (110.0113) Imaging through turbid media; (030.7060) Turbulence; (290.4210) Multiple scattering.}
  \ociscodes{(050.1755) Computational electromagnetic methods; (160.1190) Anisotropic optical materials;\\(260.1440) Birefringence; (160.3918) Metamaterials; (080.2710) Inhomogeneous optical media; (290.4210) Multiple scattering.}
  
  % To be edited by editor
  \newcommand{\thedoi}{10.1364/optica.XX.XXXXXX}
  %\doi{\url{http://dx.doi.org/\thedoi}}
}

%
%\makeatletter
%\def\@maketitle{%
%  \newpage
%  \null
%  \vskip 1em%
%  \begin{center}%
%    \let \footnote \thanks
%    {\HUGE \@title \par}%
%    \vskip 1.0em%
%    {\large
%      \lineskip .5em%
%      \begin{tabular}[t]{c}%
%        \@author
%      \end{tabular}\par}%
%    \vskip 1em%
%    {\large \@date}%
%  \end{center}%
%  \par
%  \vskip 1.0em}
%\makeatother

\makeatletter
\twocolumn[{%
  {\Huge \sffamily \bfseries \fontsize{24}{26} \selectfont \@title \par}%
  \vskip 0.5em %
  {\Huge \sffamily \fontsize{24}{26} \selectfont \@author}%
  \begin{changemargin}{1.5cm}{2.5cm}% 
    \begin{abstract}%
        Understanding the interaction of light with a highly scattering material is essential for optical microscopy of optically thick and heterogeneous biological tissues. Ensemble-averaged analytic solutions cannot provide more than general predictions for relatively simple cases. Yet, biological tissues contain chiral organic molecules and many of the cells' structures are birefringent, a property exploited by polarization microscopy for label-free imaging. Solving Maxwell's equations in such materials is a notoriously hard problem. Here we present an efficient method to determine the propagation of electro-magnetic waves in arbitrary anisotropic materials. We demonstrate how the algorithm enables large scale calculations of the scattered light field in complex birefringent materials, chiral media, and even materials with a negative refractive index.%
    \end{abstract}%
  \end{changemargin}%
  \vskip 1em %
}]
\makeatother

%\maketitle

  \section{The algorithm}\label{sec:algorithm}
    \subsection{Detailed description of the general algorithm}\label{sec:generalAlgorithm}
    	%\subsection{Description}
      The iteration loop of the algorithm is relatively short and identical for magnetic and non-magnetic materials (Algorithm~\ref{alg:generalAlgorithm}, lines~\ref{op:general_iteration_begin}-\ref{op:general_iteration_end}). Only the preconditioning steps differ. The algorithm starts by checking whether the material has magnetic properties and it determines the background permittivity, $\alpha = \alpha_\rr + \ri\alpha_\ri$, and if required, the permeability scale, $\beta$. This enables the definition of the susceptibility, $\boldsymbol{\chi}$, the dyadic Green function, $\bG$, and the source distribution, $\boldsymbol{S}$.
      
      Before starting the iteration loop, the electric field is either initialized to all zero, or to an approximate solution if available. On line~\ref{op:calc_update_general}, the loop starts by calculating the next term in the series, $\Delta\boldsymbol{E}$, using the operators $\boldsymbol{\chi}\mdot$ and $\bG\ast$ on the source distribution, $\boldsymbol{S}$, and the current estimate of the field, $\boldsymbol{E}$. The next term is added to the current estimate, $\boldsymbol{E}$, under the condition that the l$_2$-norm of the new term is less than that of the previous term. Otherwise, the series must be divergent for the current choice of the background permittivity, $\alpha$, so its imaginary part is increased by $50\%$. The iteration continues until the updates to the field are deemed sufficiently small.
      
      \begin{algorithm*}
        \caption{A function that implements the general algorithm for both non-magnetic and magnetic materials.}\label{alg:generalAlgorithm}
        \begin{algorithmic}[1]
          \Function{SolveMacroscopicMaxwell}{$\Delta\boldsymbol{r}, \boldsymbol{j}, \boldsymbol{\epsilon}, \boldsymbol{\xi}=\boldsymbol{0}, \boldsymbol{\zeta}=\boldsymbol{0}, \boldsymbol{\mu}=\mathbb{1}_3, \boldsymbol{E_0}=\boldsymbol{0},r_\mathrm{max}$}
          \If {$\boldsymbol{\xi} \equiv \boldsymbol{\zeta} \equiv \boldsymbol{0}$ and $\boldsymbol{\mu}$ is both isotropic and constant} \Comment{ non-magnetic}
          \State $\hat{\sigma}\left(\beta, \alpha_\rr\right) \equiv \left\|\boldsymbol{\epsilon} / \beta -  \mathbb{1}_3\alpha_\rr\right\|$ \Comment{ $\left\|\mdot\right\|$ is defined as the largest singular value.}
          \State $\beta \gets 1 / \boldsymbol{\mu}_{11}$
          \State $\alpha_\rr \gets \argmin{\alpha_\rr}\left[\hat{\sigma}\left(\beta, \alpha_\rr\right)\right]\;\;\;\;\forall \alpha_\rr \in \mathbb{R}$
          \Else \Comment{ magnetic}
          \State $\hat{\sigma}\left(\beta, \alpha_\rr\right) \equiv \left\|\left(\boldsymbol{\epsilon}-\boldsymbol{\xi}\mdot\boldsymbol{\mu}^{-1}\mdot\boldsymbol{\zeta}\right) / \beta -  \mathbb{1}_3\alpha_\rr\right\| + \left[\left\|\boldsymbol{\xi}\mdot\boldsymbol{\mu}^{-1}\right\| + \left\|\boldsymbol{\mu}^{-1}\mdot\boldsymbol{\zeta}\right\|\right] \sigmaCurl / \beta
          + \left\|\mathbb{1}_3 - \boldsymbol{\mu}^{-1} / \beta\right\| \sigmaCurl^2$ \LineComment{where $\sigmaCurl = k_0^{-1} \pi / \|\Delta\boldsymbol{r}\|$ is proportional to the highest possible spatial frequency}
          \State $\beta, \alpha_\rr \gets \argmin{\beta,\alpha_\rr}\left[\hat{\sigma}\left(\beta, \alpha_\rr\right)|\beta|\right]\;\;\;\;\forall \beta \in \mathbb{R}_{>0}, \alpha_\rr \in \mathbb{R}$
          \EndIf
          \State $\alpha_\ri \gets \hat{\sigma}(\beta, \alpha_\rr)$
          \State $\alpha \gets \alpha_\rr + \ri\alpha_\ri$
          \State $\boldsymbol{\chi}  \gets \frac{1}{\beta}\left(\boldsymbol{\epsilon}-\boldsymbol{\xi}\mdot\boldsymbol{\mu}^{-1}\mdot\boldsymbol{\zeta}\right) -  \mathbb{1}_3\alpha - \frac{\ri}{\beta}\boldsymbol{\xi}\mdot\boldsymbol{\mu}^{-1}\mdot\CurlOp + \frac{\ri}{\beta}\CurlOp\mdot\boldsymbol{\mu}^{-1}\mdot\boldsymbol{\zeta}
          + \CurlOp\mdot\left(\mathbb{1}_3 - \frac{1}{\beta}\boldsymbol{\mu}^{-1}\right)\mdot\CurlOp$
          
          \State $\boldsymbol{E} \gets \boldsymbol{E_0}$
          \State $p \gets \infty$ \Comment{ the l$_2$-norm of the previous update}
          \Repeat \label{op:general_iteration_begin}
          \State $\Delta\boldsymbol{E} \gets \frac{\ri}{\alpha_\ri}\boldsymbol{\chi} \mdot\left[\bG\ast \left(k_0^2\boldsymbol{\chi}\mdot\boldsymbol{E} + \boldsymbol{S}\right) - \boldsymbol{E}\right] $ \label{op:calc_update_general} \Comment{ calculate the next term in the series}
          %\State $\Delta\boldsymbol{E} \gets \mathrm{lowpass}\left(\Delta\boldsymbol{E}\right)$
          \If {$\left\|\Delta\boldsymbol{E}\right\| < p$} \label{op:divergence_check}
          \State $\boldsymbol{E} \gets \boldsymbol{E} + \Delta\boldsymbol{E}$ \Comment{ update current field estimate}
          \State $p \gets \left\|\Delta\boldsymbol{E}\right\|$
          \Else
          \State $\alpha_\ri \gets 1.5 \alpha_\ri$ \Comment{ increase $\alpha_\ri$ if divergence would be detected}
          \EndIf
          \Until{$\left\|\Delta\boldsymbol{E}\right\| < r_\mathrm{max} \left\|\boldsymbol{E}\right\|$} \label{op:general_iteration_end}
          \State \Return $\boldsymbol{E}$
          \EndFunction
        \end{algorithmic}
      \end{algorithm*}
    
      The susceptibility is defined by $\boldsymbol{\chi}  \equiv  \left(\boldsymbol{\epsilon}-\boldsymbol{\xi}\mdot\boldsymbol{\mu}^{-1}\mdot\boldsymbol{\zeta}\right) / \beta -  \mathbb{1}_3\alpha - \ri\boldsymbol{\xi}\mdot\boldsymbol{\mu}^{-1}\mdot\CurlOp / \beta + \ri\CurlOp\mdot\boldsymbol{\mu}^{-1}\mdot\boldsymbol{\zeta} / \beta
      + \CurlOp\mdot\left(\mathbb{1}_3 - \boldsymbol{\mu}^{-1} / \beta\right)\mdot\CurlOp$,
      where $\CurlOp\mdot \equiv k_0^{-1}\boldsymbol{\nabla}\times = k_0^{-1} \mathcal{F}^{-1} \mdot \bk \otimes \mathcal{F} \mdot$, with $\otimes$ the outer product, while the forward and inverse Fourier transforms are represented by $\mathcal{F}$ and $\mathcal{F}^{-1}$, respectively. Although all operations can be represented as large matrix operations, it is more space and time efficient to use fast-Fourier transforms and point-wise $3\times 3$-dot products for each application of the operator $\boldsymbol{\chi}\mdot$ on the electric field, $\boldsymbol{E}$.    
      The source is defined by $\boldsymbol{S} \equiv \ri \omega \mu_0 \; \boldsymbol{j} / \left(\beta k_0^{2}\right)$.
      The dyadic Green's function, $\bG\ast$, is discussed in section~\ref{sec:dyadicGreensFunction} in what follows.
      Note that to prevent issues with numerical precision, in the implementation the factor $k_0^{-2}$ is moved from the definition of $\bG$ to that of $\boldsymbol{S}$.
    
    \subsection{The dyadic Green's function}\label{sec:dyadicGreensFunction}
      The dyadic Green function, $\bG$, is integral to the calculation of the modified Born series:
      \begin{eqnarray}
        \boldsymbol{E} & = & \left[\sum_{\iter=0}^{\infty} \boldsymbol{M}^{\iter}\right] \mdot \boldsymbol{\Gamma}\mdot\bG\mdot\boldsymbol{S},~~\mathrm{where}\label{eqn:modified_series_supp}\\
        \boldsymbol{M} & \equiv & k_0^2\boldsymbol{\Gamma}\mdot\bG \mdot\boldsymbol{\chi}-\boldsymbol{\Gamma}+\one_3,~~\mathrm{and} \label{eqn:modified_M}\\
        \boldsymbol{\Gamma} & \equiv & \frac{\ri}{\alpha_\ri}\boldsymbol{\chi}. \label{eqn:Gamma}
      \end{eqnarray}
      Although well established in the literature of classical electromagnetism~\cite{Novotny12}, the form of the dyadic Green function that we have presented in equation~(6) of the main text may not be familiar to the reader. In this appendix we justify this equation, and show a useful representation of the Green function in terms of a unitary operator, $\boldsymbol{U}$.
    
      The vector Helmholtz equation can be separated into a homogeneous, $\Op_h$, and an inhomogeneous part, $\Op_i$, as follows
      \begin{eqnarray}
        \boldsymbol{\nabla}\times\boldsymbol{\nabla}\times\boldsymbol{E}(\bx) - k_0^2\epsilon(\bx)\boldsymbol{E}(\bx) & = & \ri\omega\mu_0\boldsymbol{j}(\bx) \nonumber	\\
        \left(\boldsymbol{\nabla}\times\boldsymbol{\nabla}\times-\alpha k_0^{2}\right)\mdot\boldsymbol{E}-k_0^{2}(\epsilon-\alpha)\mdot\boldsymbol{E} & = & \boldsymbol{S} \nonumber \\
        \left(\Op_h+\Op_i\right)\mdot\boldsymbol{E} & = & \boldsymbol{S}
      \end{eqnarray}
      The dyadic Green function, $\bG$, is defined as the impulse response solution to the homogeneous part:
      \begin{equation}
        \Op_h\mdot\bG=\boldsymbol{\nabla}\times\boldsymbol{\nabla}\times\bG(\boldsymbol{x},\boldsymbol{x}')-k_0^{2}\alpha\bG(\boldsymbol{x},\boldsymbol{x}')=\mathbb{1}_3\delta^{(3)}(\boldsymbol{x}-\boldsymbol{x}')\label{eqn:Ohinv}
      \end{equation}
            
      First, we take the Fourier transform of equation~(\ref{eqn:Ohinv}), recognizing that---due to the homogeneity of the medium---$\bG$ must be a function of $\boldsymbol{x}-\boldsymbol{x}'$, and thus of a single variable $\boldsymbol{k}$ in Fourier space
      \begin{equation}
      \bG(\bx,\bx')=\int\frac{d^{3}\boldsymbol{k}}{(2\pi)^{3}}\tilde{\boldsymbol{G}}(\boldsymbol{k}){\rm e}^{{\ri}\boldsymbol{k}\mdot(\bx-\bx')}\label{eqn:fourier_G}
      \end{equation}
      or equivalently in the operator notation used in the main text (where integrals are subsumed into our `$\mdot$' product)
      \begin{equation}
      \bG=\mathcal{F}^{-1}\mdot\tilde{\boldsymbol{G}}\mdot\mathcal{F},\label{eqn:op_F_G}
      \end{equation}
      where $\tilde{\boldsymbol{G}}$ is diagonal in Fourier space. Substituting expression (\ref{eqn:fourier_G}) into (\ref{eqn:Ohinv}) we obtain the following equation for the Fourier components of the Green function
      \begin{equation}
      \boldsymbol{k}\times\boldsymbol{k}\times\tilde{\boldsymbol{G}}(\boldsymbol{k})+k_0^{2}\alpha\tilde{\boldsymbol{G}}(\boldsymbol{k})=-\mathbb{1}_3\label{eqn:helmholtz_ft}.
      \end{equation}
      At this point we decompose the identity matrix on the right into two parts $\mathbb{1}_{3}=(\mathbb{1}_3-\boldsymbol{\Pi}_{L})+\boldsymbol{\Pi}_{L}=\boldsymbol{\Pi}_{T}+\boldsymbol{\Pi}_{L}$, where
      \begin{equation}
      \boldsymbol{\Pi}_{L}=\frac{\boldsymbol{k}\otimes\boldsymbol{k}}{\boldsymbol{k}^{2}}.
      \end{equation}
      These two operators can be physically understood as projecting out the longitudinal ($\boldsymbol{\Pi}_{L}$) and transverse ($\boldsymbol{\Pi}_{T}$) parts of the electromagnetic field, associated with electrostatic and radiative contributions respectively.  We similarly decompose the Green function as $\tilde{\boldsymbol{G}}(\boldsymbol{k})=g_L(\boldsymbol{k})\boldsymbol{\Pi}_{L}+g_{T}(\boldsymbol{k})\boldsymbol{\Pi}_{T}$, finding that equation~(\ref{eqn:helmholtz_ft}) separates into two parts
      \begin{align}
      k_0^{2}\alpha g_{L}(\boldsymbol{k})&=-1\nonumber\\
      \left(\boldsymbol{k}^{2}-\alpha k_0^{2}\right)g_{T}(\boldsymbol{k})&=1
      \end{align}
      from which we can deduce that the Green function is given by
      \begin{equation}
      \bG(\bx,\bx')=\int\frac{d^{3}\boldsymbol{k}}{(2\pi)^{3}}\left[\frac{\boldsymbol{\Pi}_{T}}{\boldsymbol{k}^{2}-\alpha k_0^{2}}-\frac{\boldsymbol{\Pi}_{L}}{\alpha k_0^{2}}\right]{\rm e}^{{\ri}\boldsymbol{k}\mdot(\bx-\bx')}
      \end{equation}
      which in our operator notation we write as
      \begin{equation}
      \bG=\mathcal{F}^{-1}\mdot\left(\frac{\boldsymbol{\Pi}_{T}}{\boldsymbol{k}^{2}-\alpha k_0^{2}}-\frac{\boldsymbol{\Pi}_{L}}{\alpha k_0^{2}}\right)\mdot\mathcal{F}\label{eqn:green_operator_supp}
      \end{equation}
      where the quantity in the rounded brackets must be understood as a matrix in both vector and Fourier indices. The dependence of the bracketed quantity on a single Fourier space variable $\boldsymbol{k}$ indicates that this matrix is diagonal in the Fourier indices, with each diagonal entry corresponding to a different value of $\boldsymbol{k}$. This completes the demonstration of equation~(\ref{eqn:green_operator_supp}).
      
      %\subsection{The dyadic Green's function as a linear combination of the identity and a unitary transformation}
      The Green function operator~(\ref{eqn:green_operator_supp}) can be written as a linear combination of the identity operator and a unitary operator, an observation that is instrumental in the study of the modified Born series~(\ref{eqn:modified_series_supp}) and its convergence. This representation of the Green function relies on the background permittivity $\alpha$ being chosen as a complex number. To find this representation we first note the following identity for any complex number $z$
      \begin{equation}
        \frac{1}{z}=\frac{1}{2{\ri}\;\Im[z]}\left(1-\frac{z^{\star}}{z}\right)=\frac{1}{2{\ri}\;\Im[z]}\left(1-\re^{-2{\ri}\phase{\,z}}\right)\label{eqn:complex_identity},
      \end{equation}
      where $\phase{\,z}$ indicates the complex argument of $z$.
      Applying identity~(\ref{eqn:complex_identity}) to (\ref{eqn:green_operator_supp}), the Green function operator (\ref{eqn:green_operator_supp}) is reduced to the following form
      \begin{equation}
        \bG=-\frac{1}{2{\ri}\alpha_\ri k_0^{2}}\left(\mathbb{1}_3-\boldsymbol{U}\right)\label{eqn:green_unitary}
      \end{equation}
      where $\alpha_\ri$ is the imaginary part of $\alpha$ and the unitary matrix $\boldsymbol{U}$ is given by
      \begin{equation}
        \boldsymbol{U}=\mathcal{F}^{-1}\mdot\left(\frac{k^{2}-\alpha^{\star}k_0^{2}}{k^{2}-\alpha k_0^{2}}\boldsymbol{\Pi}_{T}+\frac{\alpha^{\star}}{\alpha}\boldsymbol{\Pi}_{L}\right)\mdot\mathcal{F}\label{eqn:unitary_operator},
      \end{equation}
      where $k=\|\boldsymbol{k}\|$ is the $l_2$-norm of $\boldsymbol{k}$, while $\alpha^\ast$ is the complex conjugate of $\alpha$.
      The operator (\ref{eqn:unitary_operator}) is unitary, due to the fact that our Fourier transform operators can be chosen such that $\mathcal{F}^{\dagger}=\mathcal{F}^{-1}$, and thus
      \begin{multline}
        \boldsymbol{U}\mdot\boldsymbol{U}^{\dagger} =\\ \mathcal{F}^{-1}\mdot\left(\frac{k^{2}-\alpha^{\star}k_0^{2}}{k^{2}-\alpha k_0^{2}}\boldsymbol{\Pi}_{T}+\frac{\alpha^{\star}}{\alpha}\boldsymbol{\Pi}_{L}\right)\mdot
        %\mathcal{F}\mdot\mathcal{F}^{-1}\mdot
        \left(\frac{k^{2}-\alpha k_0^{2}}{k^{2}-\alpha^{\star} k_0^{2}}\boldsymbol{\Pi}_{T}+\frac{\alpha}{\alpha^{\star}}\boldsymbol{\Pi}_{L}\right)\mdot\mathcal{F}\\
        =\one_3,
      \end{multline}
      where we used the fact that $\boldsymbol{\Pi}_{T}\mdot\boldsymbol{\Pi}_{L}=0$, $\boldsymbol{\Pi}_{T}\mdot\boldsymbol{\Pi}_{T}=\boldsymbol{\Pi}_{T}$, and $\boldsymbol{\Pi}_{L}\mdot\boldsymbol{\Pi}_{L}=\boldsymbol{\Pi}_{L}$.  
      
      Note that the unitary operator only changes the complex argument of the transverse and longitudinal projected components.
      It can thus be seen from equation~(\ref{eqn:green_unitary}) that the eigenvalue of $\bG$ with the largest absolute value equals $\frac{\ri}{\alpha_\ri k_0^2}$, and that the real part of the eigenvalue must be no larger than $\frac{1}{2 \alpha_\ri k_0^2}$ in absolute value, while its imaginary part cannot be negative. More specifically, all eigenvalues are contained within a disk of radius $\frac{1}{2 \alpha_\ri k_0^2}$ around the point $\frac{\ri}{2 \alpha_\ri k_0^2}$ in the complex plane. %\todo[inline]{TV: This last sentence is only necessary for the geometrical discussion}
      
      We note that for numerical purposes it is space and time-efficient to implement the dyadic Green function operation in k-space. The resulting multiplication only requires space to store the Fourier-transformed input, output, and Green function. The vector operations of the latter can be implemented without constructing the full, multi-dimensional, matrix for the dyadic Green function, though at least a scalar array,  $|\bk|^2$, for normalizing the k-vectors must be stored.
    
    \subsection{Computation and memory efficiency}\label{sec:efficiency}
      The computational efficiency of the modified Born series iteration has been discussed previously by Osnabrugge~et~al.~\cite{Osnabrugge16}. In general the convergence of the iteration is approximately inversely proportional to the range of susceptibilities in the calculation volume. The method is therefore not efficient for calculations that involve metals. The anisotropic algorithm is no different in this respect. It is most efficient for heterogeneous dielectric materials such as biological tissue. However, the anisotropic version does require the addition of $3$-vectors and multiplications of $3\times 3$ matrices instead of scalar operations. The algorithm for anisotropic permittivity can thus be expected to be $9$ times slower than the scalar wave algorithm~\cite{Osnabrugge16}, and $3$ times slower than the isotropic vector algorithm~\cite{Krueger17}. It should also be noted that simulating inhomogeneous magnetic properties introduces the discretized differential operator, $\CurlOp$. This largest singular value of this operator depends on the sampling density of the computation volume. This is equivalent to having a large variation in optical properties in the sample, which is known to lead to significantly slower convergence of the modified Born series~\cite{Osnabrugge16}.   
    
      The main limitation of any large scale electro-magnetic calculation is computer memory. The advantage of the presented algorithm is that the required memory scales with the number of sample points, $P$. This is important, considering that the calculation volumes of interest can be several orders of magnitude larger than the wavelength in all three dimensions. At a very minimum the electro-magnetic field in the calculation volume must be stored. The magnetic field can be calculated from the electric field, hence it is sufficient to store only the electric vector field, $\boldsymbol{E}$. This can be done using $3P$ complex floating point numbers to represent a single frequency field in the calculation volume. Each iteration calculates a correction term for the field, $\Delta\boldsymbol{E}$, thus bringing the total to $6P$.
      
      Anisotropic permittivity can be represented by a $3\times 3$ matrix for each point in space, to make up a block-diagonal matrix of dimension $3P\times3P$. However, this can be efficiently stored using $9P$ complex numbers, while all matrix operations, as well as the fast-Fourier transforms can be performed in-place. It is not necessary to explicitly calculate the matrix for the dyadic Green's function (\ref{sec:dyadicGreensFunction}), though it is necessary to determine the normalization factor $\|\bk\|$ for all k-vectors. This requires storage for $P$ values.
      
      Anisotropy in the permeability, $\boldsymbol{\mu}$, or the coupling factors, $\boldsymbol{\xi}$ and $\boldsymbol{\zeta}$, will increase the memory requirements accordingly.    
      If the source current $\boldsymbol{S}$ is not sparse, a further $3P$ complex values need to be stored. The memory requirements are summarized in Table~\ref{tab:storage_requirements} for the case of a spatially variant $\boldsymbol{\epsilon}$, and optionally, a spatially variant $\boldsymbol{\mu}$, $\boldsymbol{\xi}$, and $\boldsymbol{\zeta}$.    
      The memory requirements are listed in different columns for the isotropic and the anisotropic cases. Mixed cases are omitted for brevity.
      
      \begin{table}[htbp]
        \centering
        \caption{\bf Storage requirements for the algorithm.}
        \begin{tabular}{ccc}
          \hline
          spatial variability & isotropic & anisotropic \\
          \hline
          $\boldsymbol{\epsilon}$ & $11P$ & $19P$ \\
          $\boldsymbol{\epsilon}$, and $\boldsymbol{\mu}$ & $12P$ & $28P$ \\
          $\boldsymbol{\epsilon}$, $\boldsymbol{\mu}$, $\boldsymbol{\xi}$, and $\boldsymbol{\zeta}$ & $14P$ & $46P$ \\
          \hline
        \end{tabular}
        \label{tab:storage_requirements}
      \end{table}

    \subsection{Sampling and prevention of aliasing}\label{sec:sampling}
      The sampling grid size and density are important considerations for the calculation accuracy. In Algorithm~\ref{alg:generalAlgorithm}, every iteration requires a multiplication by the preconditioned general susceptibility, $\boldsymbol{\chi}$, an addition of the source, $\boldsymbol{S}$, a convolution with the dyadic Green's function, and another multiplication by $\boldsymbol{\chi}$. Multiplications and additions cannot increase the spatial extent of the field $\boldsymbol{E}$ beyond the union of that of its left and right hand sides. However, the Green function has an infinite extent and therefore also the result of its convolution with a finite field. When the convolution operation is implemented as a multiplication in Fourier space, periodic boundary conditions are implicitly assumed. Alternative boundary conditions can be readily simulated by defining layers of non-transmitting material at the boundary. The sampling grid size must therefore be sufficiently large to fit both the volume of interest and multi-cell boundaries that adequately absorb the field.
      
      To discuss the sampling density, we consider the Fourier transform of the iteration update calculation:
      \begin{eqnarray}
        \boldsymbol{\Delta\widetilde{\boldsymbol{E}}} & = & \frac{\ri}{\alpha_\ri}\widetilde{\boldsymbol{\chi}}\ast\left[\widetilde{\boldsymbol{G}}\mdot\left(\widetilde{\boldsymbol{\chi}}\ast\widetilde{\boldsymbol{E}} + \widetilde{\boldsymbol{S}}\right) - \widetilde{\boldsymbol{E}} \right].
      \end{eqnarray}
      In Fourier space, each step requires a convolution, $\ast$, with the Fourier transform of the susceptibility, $\widetilde{\boldsymbol{\chi}}$, followed by an addition with the Fourier transform of the source, $\widetilde{\boldsymbol{S}}$, a multiplication with the Fourier transform of the dyadic Green's function, $\widetilde{\boldsymbol{G}}$, and another convolution with $\widetilde{\boldsymbol{\chi}}$. Although the multiplication by $\widetilde{\boldsymbol{G}}$, tends to suppress high spatial frequencies, it does not strictly limit the bandwidth support of the product. On the other hand, convolutions with $\widetilde{\boldsymbol{\chi}}$ do extend the bandwidth support twice per iteration. If we define $W_{\boldsymbol{\chi}}$, $W_{\boldsymbol{E}}$, and $W_{\boldsymbol{S}} \leq W_{\boldsymbol{\chi}}+W_{\boldsymbol{E}}$ as the spatial-frequency band-widths of $\boldsymbol{\chi}$, $\boldsymbol{E}$, and $\boldsymbol{S}$, respectively; it can seen that the bandwidth of the iteration update, $\boldsymbol{\Delta\boldsymbol{E}}$, must be $W_{\boldsymbol{\Delta\boldsymbol{E}}} \leq 2W_{\boldsymbol{\boldsymbol{\chi}}} + W_{\boldsymbol{\boldsymbol{E}}}$. Therefore, even when both the material properties and the source are smooth functions with a finite bandwidth, the sampling density of the calculation must, in principle, be increased by $2W_{\boldsymbol{\chi}}$ with every iteration step.    
      
      In practice, the calculation must be performed on a sample grid with finite bandwidth, $W$, and sample spacing, $W^{-1}$. With the notable exception of superoscillations, as long as the material properties and source field are spatially band-limited, the solution for the electric field can be expected to be concentrated around the Ewald sphere. We found that the smoothing effect of the Green function is generally sufficient to suppress the highest spatial frequencies that are affected by aliasing. When we accept the approximation that the solution must be band-limited, aliasing can be completely eliminated by low-pass filtering the field after each iteration step. To achieve this, the calculation must be performed with a sampling band-width $W \geq W_{\boldsymbol{\chi}} + W_{\boldsymbol{E}}$, in other words, with a sampling density that is no smaller than the sum of the Nyquist rates for $\boldsymbol{E}$ and $\boldsymbol{\chi}$. The two convolutions with $\widetilde{\boldsymbol{\chi}}$ expand the support in frequency space to $2W_{\boldsymbol{\chi}} + W_{\boldsymbol{E}}$, thereby causing aliasing in the upper $W_{\boldsymbol{\chi}}$-part of the band. However, the lowest spatial frequencies in the $W_{\boldsymbol{E}}$-band are not affected. All aliasing artefacts can thus be avoided by eliminating all but the spatial frequencies within the lower $W_{\boldsymbol{E}}$-band of the iteration update, $\boldsymbol{\Delta\boldsymbol{E}}$.
      Since the suppression of the highest spatial frequencies can be applied at the end of each iteration as a projection onto a subspace, it can be seen that convergence must also hold in this sub-space.
            
      It should also be noted that the algorithm as it is presented here requires the material properties to be sampled on a regular grid. This enables efficient convolutions using the fast-Fourier transform and it simplifies the implementation in general. However, one could imagine structures that require a higher sampling density in specific regions. To address such need, we envision that the method can be extended to irregular grids by using non-uniform Fourier transforms to perform the convolutions~\cite{Dutt93,Fessler03}.
                    
    \subsection{Robustness to errors}\label{sec:robustness_to_errors}
      Unlike time-stepped methods such as FDTD, the iteration presented here is robust to numerical errors.
      The $\iterend^\mathrm{th}$-correction term, $\Delta\boldsymbol{E}_\iterend$, is obtained from recursive equation~(\ref{eqn:modified_series_supp}):
      \begin{eqnarray}
      \boldsymbol{E}_\iterend & = & \boldsymbol{\Gamma}\mdot\bG\ast k_0^2\boldsymbol{\chi}\mdot\boldsymbol{E}_{\iterend-1} - \boldsymbol{\Gamma}\mdot\boldsymbol{E}_{\iterend-1} + \boldsymbol{E}_{\iterend-1} + \boldsymbol{\Gamma}\mdot\bG\ast\boldsymbol{S}\nonumber\\
      \Rightarrow \Delta\boldsymbol{E}_\iterend & = & \boldsymbol{E}_\iterend - \boldsymbol{E}_{\iterend-1} =  \boldsymbol{\Gamma}\mdot\left[\bG\ast\left(k_0^2\boldsymbol{\chi}\mdot\boldsymbol{E}_{\iterend-1}+\boldsymbol{S}\right)-\boldsymbol{E}_{\iterend-1}\right].\nonumber
      \end{eqnarray}
      For the initial field estimate, 
      If we begin with a null field estimate $\boldsymbol{E}_{0} = \boldsymbol{0}$, this iteration is equivalent to the modified series~(\ref{eqn:modified_series_supp}). However, it should be noted that when $\boldsymbol{E}_{0} \neq \boldsymbol{0}$, the iteration corresponds to a different series, yet one that converges under the same conditions and to the same limit:
      \begin{eqnarray}
      \boldsymbol{E}_\iterend & = & \boldsymbol{\Gamma}\mdot\left[\bG\ast\left(k_0^2\boldsymbol{\chi}\mdot\boldsymbol{E}_{\iterend-1} + \boldsymbol{S}\right) - \boldsymbol{E}_{\iterend-1}\right] + \boldsymbol{E}_{\iterend-1}\nonumber\\
      & = & \boldsymbol{M}\mdot\boldsymbol{E}_{\iterend-1} + \boldsymbol{\Gamma}\mdot\bG\ast\boldsymbol{S}\nonumber\\
      & = & \boldsymbol{M}^\iterend\mdot\boldsymbol{E}_{0} + \left[\sum_{\iter=0}^{\iterend-1}\boldsymbol{M}^{\iter}\right]\mdot\boldsymbol{\Gamma}\mdot\bG\ast\boldsymbol{S}
      \end{eqnarray}
      It can be seen that the $\iterend^\mathrm{th}$-term of the series differs by $\boldsymbol{M}^N\mdot\boldsymbol{E}_{0}$ from that of series~(\ref{eqn:modified_series_supp}), and that its corresponding residue, $\left\|\boldsymbol{E}-\boldsymbol{E}_\iterend\right\|$, is
      \begin{multline} \left\|\left[\sum_{\iter=0}^\infty\boldsymbol{M}^{\iter}\right]\mdot\boldsymbol{\Gamma}\mdot\bG\ast\boldsymbol{S} - \boldsymbol{M}^\iterend\mdot\boldsymbol{E}_0 - \left[\sum_{\iter=0}^{\iterend-1}\boldsymbol{M}^{\iter}\right]\mdot\boldsymbol{\Gamma}\mdot\bG\ast\boldsymbol{S}\right\| \\
      = \left\|\boldsymbol{M}^\iterend\mdot\left(\boldsymbol{E} - \boldsymbol{E}_0\right)\right\|.\label{eqn:residue}
      \end{multline}
      Provided that the absolute largest eigenvalue of $\boldsymbol{M}$ is smaller than one, the upper bound on the residue is tightened with every iteration, independently of the choice of the initial field. However, it can be seen from equation~(\ref{eqn:residue}) that starting from an approximate solution has a similar effect as skipping the first iterations, and allowing the algorithm to reach convergence in less time. The corrections in consecutive terms also prevent the accumulation of numerical errors, a common issue with techniques such as the finite-difference time-domain method.
  
      Since the iteration does not have to be started from the all-zero field, convergence may be reached faster if an approximate solution is provided as the initial field of the iteration. It is straightforward to start the algorithm with an approximate solution to the corresponding isotropic or scalar problem. Techniques such as the beam-propagation method or two-dimensional fast marching~\cite{Capozzoli14}, could be leveraged as an initialization of the anisotropic algorithm so to reduce the total run-time.

%
%: Convergence of modified Born series
%
\section{Convergence of the modified Born series}\label{sec:proof}
  	\subsection{Geometrical Interpretation}
      The iteration as given by equation~(\ref{eqn:modified_series_supp}) calculates a series of correction terms of the form $\boldsymbol{M}^\iter\mdot\boldsymbol{E}_0$. Independently of the initial value $\boldsymbol{E}_0$, this series is guaranteed to converge when all eigenvalues of $\boldsymbol{M}$ are less than one in absolute value. Recall that the preconditioned iteration operator, $\boldsymbol{M}$, is given by definition~(\ref{eqn:modified_M}).
      
  		In the previous section it is shown with equation~(\ref{eqn:green_unitary}) that the dyadic Green function, $\bG$, can be decomposed in the scaled sum of an identity and a unitary transformation, $\boldsymbol{U}$:
  		\begin{eqnarray}
			\bG & \equiv & \frac{\one_3 - \boldsymbol{U}}{2} \frac{\ri}{\alpha_\ri k_0^2},
  		\end{eqnarray}
      Substitution in equation~(\ref{eqn:modified_M}) allows it to written in terms of just the preconditioner, $\boldsymbol{\Gamma}$, and the susceptibility, $\boldsymbol{\chi}$:
  		\begin{eqnarray}
	  		\boldsymbol{M} & \equiv & \Gamma\mdot\left(\frac{\one_3 - \boldsymbol{U}}{2}\mdot\frac{\ri}{\alpha_\ri}\boldsymbol{\chi}-\one_3\right) + \one_3.\label{eqn:geometricalConstructionM}
	  	\end{eqnarray}
    	The operator $\boldsymbol{\Gamma}$ must be chosen so that the largest eigenvalue of the preconditioned iteration operator, $\boldsymbol{M}$, is less than $1$ in absolute value. In other words, $\left|\boldsymbol{E}_\maxeig^\dagger\mdot\boldsymbol{M}\mdot\boldsymbol{E}_\maxeig\right| < 1$, where $\boldsymbol{E}_\maxeig$ is the eigenvector of the largest eigenfunction of $\boldsymbol{M}$. We will now investigate the effects of these operations using a geometrical construction of the complex values that $\boldsymbol{E}_\maxeig^\dagger\mdot\boldsymbol{M}\mdot\boldsymbol{E}_\maxeig$ can take and show under what conditions these fall within the unit disk around the origin in the complex plane.
	  	
	  	\begin{figure}[ht!]
	  		\centering
	  		\fbox{
	  			\includegraphics[width=0.97\linewidth]{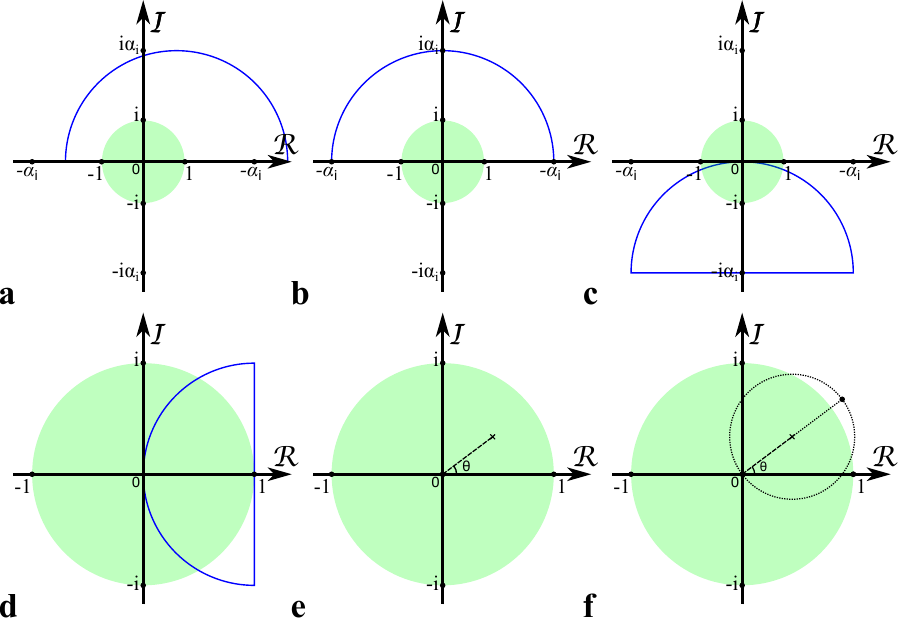}
	  		}
    		\caption{Geometrical interpretation of the multiplication by the susceptibility,  $\boldsymbol{\chi}$, and the dyadic Green function, $\bG$. Values are shown in the complex plane and the final values must be within the unit disk (green) to ensure convergence. The blue semi-circle indicates the possible values of $\boldsymbol{E}_\maxeig^\dagger\mdot\boldsymbol{\epsilon}\mdot\boldsymbol{E}_\maxeig$ (a), $\boldsymbol{E}_\maxeig^\dagger\mdot\left(\boldsymbol{\epsilon}-\alpha_\rr\one_3\right)\mdot\boldsymbol{E}_\maxeig$ (b), $\boldsymbol{E}_\maxeig^\dagger\mdot\boldsymbol{\chi}\mdot\boldsymbol{E}_\maxeig$ (c), and , $\boldsymbol{E}_\maxeig^\dagger\mdot\frac{\ri}{\alpha_\ri}\boldsymbol{\chi}\mdot\boldsymbol{E}_\maxeig$ (d), associated with the susceptibility. (e, f) The dotted circles with dashed radius represents the decomposition of the dyadic Green function in the sum of an identity (dashed line between origin and $\times$) and unitary operation (dotted line and circle around point marked with $\times$).}
	  		\label{fig:geometricalProofIntro}
	  	\end{figure}
	  	  	
    	The expression $\boldsymbol{E}_\maxeig^\dagger\mdot\boldsymbol{M}\mdot\boldsymbol{E}_\maxeig$ is constructed step-by-step, starting from the permittivity, $\boldsymbol{\epsilon} \equiv \boldsymbol{\chi} + \alpha\one_3$. The real and imaginary parts of $\boldsymbol{E}_\maxeig^\dagger\mdot\boldsymbol{\epsilon}\mdot\boldsymbol{E}_\maxeig$ can be seen to be:
      \begin{eqnarray}
        \Re\left\{\boldsymbol{E}_\maxeig^\dagger\mdot\boldsymbol{\epsilon}\mdot\boldsymbol{E}_\maxeig\right\} & = & \boldsymbol{E}_\maxeig^\dagger\mdot\frac{\boldsymbol{\epsilon} + \boldsymbol{\epsilon}^\dagger}{2}\mdot\boldsymbol{E}_\maxeig\;\; \mathrm{and}\nonumber\\
        \Im\left\{\boldsymbol{E}_\maxeig^\dagger\mdot\boldsymbol{\epsilon}\mdot\boldsymbol{E}_\maxeig\right\} & = & \boldsymbol{E}_\maxeig^\dagger\mdot\frac{\boldsymbol{\epsilon} - \boldsymbol{\epsilon}^\dagger}{2\ri}\mdot\boldsymbol{E}_\maxeig, 
      \end{eqnarray}
      where gain-free media have a positive definite imaginary part. Assuming that the medium is free from gain and finite-valued, then there must exist a semi-circle as shown in blue in Figure~\ref{fig:geometricalProofIntro}(a) that contains all values $\boldsymbol{E}_\maxeig^\dagger\mdot\boldsymbol{\epsilon}\mdot\boldsymbol{E}_\maxeig$ in the upper half of the complex plane. The values $\boldsymbol{E}_\maxeig^\dagger\mdot\left(\boldsymbol{\epsilon} - \alpha_\rr\one_3\right)\mdot\boldsymbol{E}_\maxeig$ are contained within the same semi-circle, now centered at the origin as shown in Figure~\ref{fig:geometricalProofIntro}(b). Although $\boldsymbol{E}_\maxeig$ is not necessarily also an eigenfunction of $\boldsymbol{\chi}$, it can be seen that when $\boldsymbol{\epsilon}$ is positive definite, such semi-circle must exist with radius, $\alpha_\ri$, equal to the largest singular value of $\boldsymbol{\epsilon} - \alpha_\rr\one_3$. The values $\boldsymbol{E}_\maxeig^\dagger\mdot\left(\boldsymbol{\epsilon} - \alpha\one_3\right)\mdot\boldsymbol{E}_\maxeig = \boldsymbol{E}_\maxeig^\dagger\mdot\boldsymbol{\chi}\mdot\boldsymbol{E}_\maxeig$ can be seen to be contained within the semi-circle centered at $-\ri\alpha_\ri$ and with its circumference touching the origin, as shown in Figure~\ref{fig:geometricalProofIntro}(c). The multiplication by $\frac{\ri}{\alpha_\ri}$ normalizes and rotates the semi-circle $90^\circ$ around the origin for the values $\boldsymbol{E}_\maxeig^\dagger\mdot\frac{\ri}{\alpha_\ri}\boldsymbol{\chi}\mdot\boldsymbol{E}_\maxeig$ as shown in Figure~\ref{fig:geometricalProofIntro}(d). The radius of the semi-circle is now $1$, and its base diagonal parallel to the imaginary axis through the point $1+0\ri$ on the real axis.
	  	
      When the dyadic Green function tensor is applied to a point in Hilbert space, the result is contained within the sphere with as diagonal the line-segment between the origin and the original point, scaled by $\frac{\ri}{\alpha_\ri k_0^2}$. The scaling factor has already been treated together with the susceptibility. To understand the effect of the dyadic Green function, we decompose $\boldsymbol{E}_\maxeig$ in the eigenfunction basis of $\frac{\ri}{\alpha_\ri}\boldsymbol{\chi}$ as $\boldsymbol{E}_\maxeig = \sum_i c_i\boldsymbol{E}_i$, and first consider its effect on the eigenfunctions $\boldsymbol{E}_i$ of $\frac{\ri}{\alpha_\ri}\boldsymbol{\chi}$ with eigenvalues $\lambda_i$. Figure~\ref{fig:geometricalProofIntro}(e) shows how the circle's center point is constructed by the identity term in the dyadic Green function $\boldsymbol{E}_i^\dagger\mdot\frac{\one_3}{2}\mdot\left(\frac{\ri}{\alpha_\ri}\boldsymbol{\chi}\mdot\boldsymbol{E}_i\right)$, while the inclusion of the unitary term, $\boldsymbol{E}_i^\dagger\mdot\frac{\one_3 - \boldsymbol{U}}{2}\mdot\left(\frac{\ri}{\alpha_\ri}\boldsymbol{\chi}\mdot\boldsymbol{E}_i\right)$, outlines the circle's circumference as shown in Figure~\ref{fig:geometricalProofIntro}(f). Note that the unitary transformation may reduce the magnitude of the final projection, hence the circle only indicates the bounds of a disk of possible values. 
      
	  	\begin{figure}[ht!]
	  		\centering
	  		\fbox{
	  			\includegraphics[width=0.97\linewidth]{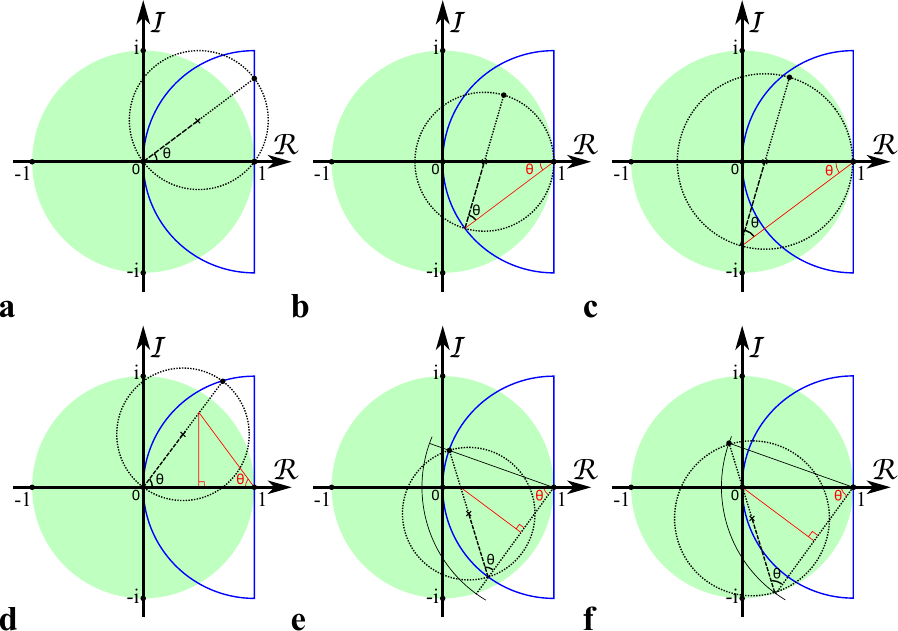}
	  		}
	  		\caption{Geometrical interpretation of the converging iteration. The possible values that the expression $\boldsymbol{E}_i^\dagger\mdot\boldsymbol{M}\mdot\boldsymbol{E}_i$ can take are constructed in the complex plane. The susceptibilities can be separated into two limiting cases: purely reactive susceptibility without absorption represented by the diagonal of the semi-circle (a-c), and those corresponding to a permittivity with an absolute value $\alpha_\ri$ on the circumference (d-f). The red lines indicate geometrical constructions of equilateral triangles. (a,d) The values for $\boldsymbol{E}_i^\dagger\mdot\frac{\one_3 - \boldsymbol{U}}{2}\mdot\frac{\ri}{\alpha_\ri}\boldsymbol{\chi}\mdot\boldsymbol{E}_i$ must lie within the black dotted circle with as diagonal the line segment between the origin and any point in the blue semi-circle. The preconditioner $\boldsymbol{\Gamma}$ rotates (b,e) and scales (c,f) the values around the point $1+0\ri$ on the real axis.}
	  		\label{fig:geometricalProof}
	  	\end{figure}
	  	
      The combined effects of the susceptibility and the dyadic Green function allow us to study the bounds on the eigenvalues of the preconditioned iteration operation $\boldsymbol{M}$. The value of $\boldsymbol{E}_i^\dagger\mdot\frac{\one_3 - \boldsymbol{U}}{2}\frac{\ri}{\alpha_\ri}\boldsymbol{\chi}\mdot\boldsymbol{E}_i$ can be inside any disk between the origin and any point within the previously described semi-circle. Figure~\ref{fig:geometricalProof}(a) shows the semi-circle in blue and the radii and circumference of a disk for one specific example in black dotted line. It can be noted that some values of $\boldsymbol{E}_i^\dagger\mdot\frac{\one_3 - \boldsymbol{U}}{2}\frac{\ri}{\alpha_\ri}\boldsymbol{\chi}\mdot\boldsymbol{E}_i$ still fall outwith the unit circle in the complex plane, marked as a green background. This lies at the basis of the divergence and emphasizes the need for the preconditioner. The final three operations to construct $\boldsymbol{E}_i^\dagger\mdot\boldsymbol{M}\mdot\boldsymbol{E}_i$ as defined in equation~(\ref{eqn:geometricalConstructionM}) are a subtraction of the identity, a multiplication by the preconditioner $\boldsymbol{\Gamma}$, and the re-addition of the identity. Since $\boldsymbol{\Gamma}$ is a linear operation, it can be understood as a scaling and rotation around the origin in the complex plane, and the three combined operations are the same scaling and rotation, but around the point $1+0\ri$. We will now show that this ensures that all values fall within the unit disk. The black dotted circle shown in Figure~\ref{fig:geometricalProof}(a) goes through the point $1+0\ri$ when $\boldsymbol{E}_i^\dagger\mdot\frac{\ri}{\alpha_\ri}\boldsymbol{\chi}\mdot\boldsymbol{E}_i$ coincides with the base of the blue semi-circle because the base and the real axis form a right-angle triangle with the diagonal of the black dotted circle. To ensure that all points of the black dotted circle shown in Figure~\ref{fig:geometricalProof}(a) are on the unit disk, one has to rotate it around the point $1+0\ri$ and make it tangent to the unit disk. It can be seen that this can only be achieved when the operation $\boldsymbol{\Gamma}$ introduces a phase $\theta$ that is equal to that introduced by $\frac{\ri}{\alpha_\ri}\boldsymbol{\chi}$ in the previous step. In other words, $\theta$ is the complex argument of the previously studied expression $\boldsymbol{E}_i^\dagger\mdot\frac{\ri}{\alpha_\ri}\boldsymbol{\chi}\mdot\boldsymbol{E}_i$. This operation rotates the origin of the black dotted circle onto the real axis. The multiplication by $\boldsymbol{\Gamma}$ can also scale the result. The scaling factor should be positive and sufficiently small so that the resulting radius remains below one. When the same scaling factor as $\boldsymbol{E}_i^\dagger\mdot\frac{\ri}{\alpha_\ri}\boldsymbol{\chi}\mdot\boldsymbol{E}_i$ is used, the equilateral triangle is scaled to touch the imaginary axis as shown in Figure~\ref{fig:geometricalProof}(c). The base of equilateral triangle (shown in red), can be seen to trace out the values $\boldsymbol{E}_i^\dagger\mdot\left(\one_3 - \frac{\ri}{\alpha_\ri}\boldsymbol{\chi}\right)\mdot\boldsymbol{E}_i$ inside the unit disk. Although a larger scaling factor would be permissible in the particular case shown in Figure~\ref{fig:geometricalProofIntro}(c), it is convenient to define the preconditioner as $\boldsymbol{\Gamma}\equiv\frac{\ri}{\alpha_\ri}\boldsymbol{\chi}$, to ensure that all the eigenvalues are contained within the unit disk.
	  	
      Similarly one can see that the values $\boldsymbol{E}_i^\dagger\mdot\frac{\one_3 - \boldsymbol{U}}{2}\frac{\ri}{\alpha_\ri}\boldsymbol{\chi}\mdot\boldsymbol{E}_i$ on the circumference of the blue semi-circle are mapped to fall into the unit disk when using the same rotation angle, $\theta$, and scaling factor as $\boldsymbol{E}_i^\dagger\mdot\frac{\ri}{\alpha_\ri}\boldsymbol{\chi}\mdot\boldsymbol{E}_i$. Figure~\ref{fig:geometricalProof}(d) shows how a different equilateral triangle can be constructed between the diagonal of the black dotted circle and the real axis. When the rotation caused by $\boldsymbol{\Gamma}$ is equal to that in the previous step, it rotates the apex of the equilateral triangle onto the real axis as shown in Figure~\ref{fig:geometricalProof}(e). The black dotted circle now falls within the bounds of the unit disk. When the preconditioner $\boldsymbol{\Gamma}$ scales the result as in the previous step, the apex translates to coincide with the origin and the diagonal of the black dotted circle goes through the origin. The points connecting the base of the equilateral triangle must therefore lie on the same circle. Since one of the endpoints lies on the circumference of the unit disk, the diagonal line of the black dotted circle must fall entirely within the unit disk.

      Values that are in between the two extreme cases discussed above can be written as a weighted sum. The linearity of the operation and the convexity of the unit sphere guarantees that also non-extreme values have eigenvalues that are no greater than those encountered at the boundaries. While this shows that the $\left|\boldsymbol{E}_i^\dagger\mdot\boldsymbol{M}\mdot\boldsymbol{E}_i\right| \leq 1$ for all eigenfunctions, $\boldsymbol{E}_i$, of $\boldsymbol{\chi}$; the same is not necessarily true for, $\boldsymbol{E}_\maxeig$, an eigenfunction of $\boldsymbol{M}$ that may be a linear combinations of the eigenfunctions $\boldsymbol{E}_i$ of $\boldsymbol{\chi}$. 

      The general expression $\boldsymbol{E}_\maxeig^\dagger\mdot\boldsymbol{M}\mdot\boldsymbol{E}_\maxeig$ can be written as the difference
      \begin{multline}
        \boldsymbol{E}_\maxeig^\dagger\mdot\left[\boldsymbol{\Gamma}\mdot\left(\frac{\one_3 - \boldsymbol{U}}{2}\mdot\frac{\ri}{\alpha_\ri}\boldsymbol{\chi} - \one_3 \right)+ \one_3\right]\mdot\boldsymbol{E}_\maxeig=\\
        \boldsymbol{E}_\maxeig^\dagger\mdot\left[\boldsymbol{\Gamma}\mdot\left(\frac{1}{2}\frac{\ri}{\alpha_\ri}\boldsymbol{\chi} - \one_3 \right)+ \one_3\right]\mdot\boldsymbol{E}_\maxeig - \boldsymbol{E}_\maxeig^\dagger\mdot\boldsymbol{\Gamma}\mdot\frac{\boldsymbol{U}}{2}\mdot\frac{\ri}{\alpha_\ri}\boldsymbol{\chi}\mdot\boldsymbol{E}_\maxeig.
      \end{multline}
      In the special case that the eigenfunctions of $\boldsymbol{\chi}$ are orthonormal and using $\bG\equiv\frac{\ri}{\alpha_\ri}\boldsymbol{\chi}$, the first term can be rewritten as a weighted sum over the complex values corresponding to each eigenfunction of the system using $\boldsymbol{E}_{\max}=\sum_i c_i \boldsymbol{E}_i$ with $c_i = \boldsymbol{E}_{i}^\dagger \mdot \boldsymbol{E}_{\max}$ and  $\sum_i |c_i|^2 = 1$:
      \begin{multline}
        \boldsymbol{E}_\maxeig^\dagger\mdot\boldsymbol{M}\mdot\boldsymbol{E}_\maxeig =\\
         \sum_i |c_i|^2 \left[\frac{\lambda_i^2}{2} - \lambda_i + 1\right] - \frac{1}{2}\left(\boldsymbol{\Gamma}^\dagger\mdot\boldsymbol{E}_\maxeig\right)^\dagger\mdot\boldsymbol{U}\mdot\frac{\ri}{\alpha_\ri}\boldsymbol{\chi}\mdot\boldsymbol{E}_\maxeig.\label{eqn:geometricalProofLollypopDifference}
      \end{multline}
      Here, the summation term defines the center position of the circle for $\boldsymbol{E}_\maxeig$ as a weighted average of the center positions of the circles corresponding to the eigenfunctions $\boldsymbol{E}_i$. Since for each eigenfunction, a circle with radius $\frac{1}{2}|\lambda_i|^2$ fits within the unit disk of the complex plane so the absolute value of the summation term is limited by the inequality
      \begin{multline}
        \left| \sum_i |c_i|^2 \left[\frac{\lambda_i^2}{2} - \lambda_i + 1\right]\right| \leq \sum_i |c_i|^2\left[1 - \frac{|\lambda_i|^2}{2}\right]
      =\\
      \sum_i |c_i|^2 - \sum_i |c_i|^2\frac{|\lambda_i|^2}{2} = 1 - \frac{1}{2}\sum_i |c_i|^2|\lambda_i|^2.
      \end{multline}
      This leaves sufficient space to fit a circle of radius $\frac{1}{2}\sum_i |c_i|^2 |\lambda_i|^2$ within the unit disk of the complex plane. The second term in equation~(\ref{eqn:geometricalProofLollypopDifference}) is one-half of the dot product of a unitary operation and two terms with identical $l_2$-norm $\left\|\frac{\ri}{\alpha_\ri}\boldsymbol{\chi}\mdot\boldsymbol{E}_\maxeig\right\| = \sqrt{\sum_i |c_i|^2 |\lambda_i|^2} = \left\|\boldsymbol{\Gamma}^\dagger\mdot\boldsymbol{E}_\maxeig\right\|$, when $\boldsymbol{\Gamma}$ is defined as $\frac{\ri}{\alpha_\ri}\boldsymbol{\chi}$. Hence, by defining $\boldsymbol{\Gamma} \equiv \frac{\ri}{\alpha_\ri}\boldsymbol{\chi}$, it can be seen that equation~(\ref{eqn:geometricalProofLollypopDifference}) cannot be larger than $1$, where $\alpha_\ri$ is larger than the largest singular value of $\|\bD\|$. The series must thus converge when the eigenfunctions of the susceptibility distribution, $\boldsymbol{\chi}$, are orthogonal, a very common situation. Yet, its eigenfunctions will not be orthogonal when the reactive, $\bD_\rr$, and dissipative parts, $\bD_\ri$, of $\bD=\boldsymbol{\chi}+{\ri}\alpha_\ri\one_3 = \bD_\rr +\ri\bD_\ri$ do not commute. This would occur when a birefringent crystal also has a polarization dependent absorption, yet with a different axis. In what follows, such more general susceptibilities are analyzed.

      %
      %: Numerical experiments
      %
      \subsection{Numerical demonstration of the convergence}\label{sec:numerical_demonstration}
        %
        %: Numerical experiments figure
        %
        \begin{figure}[ht!]
          \centering
            \fbox{
            	\includegraphics[width=0.97\linewidth]{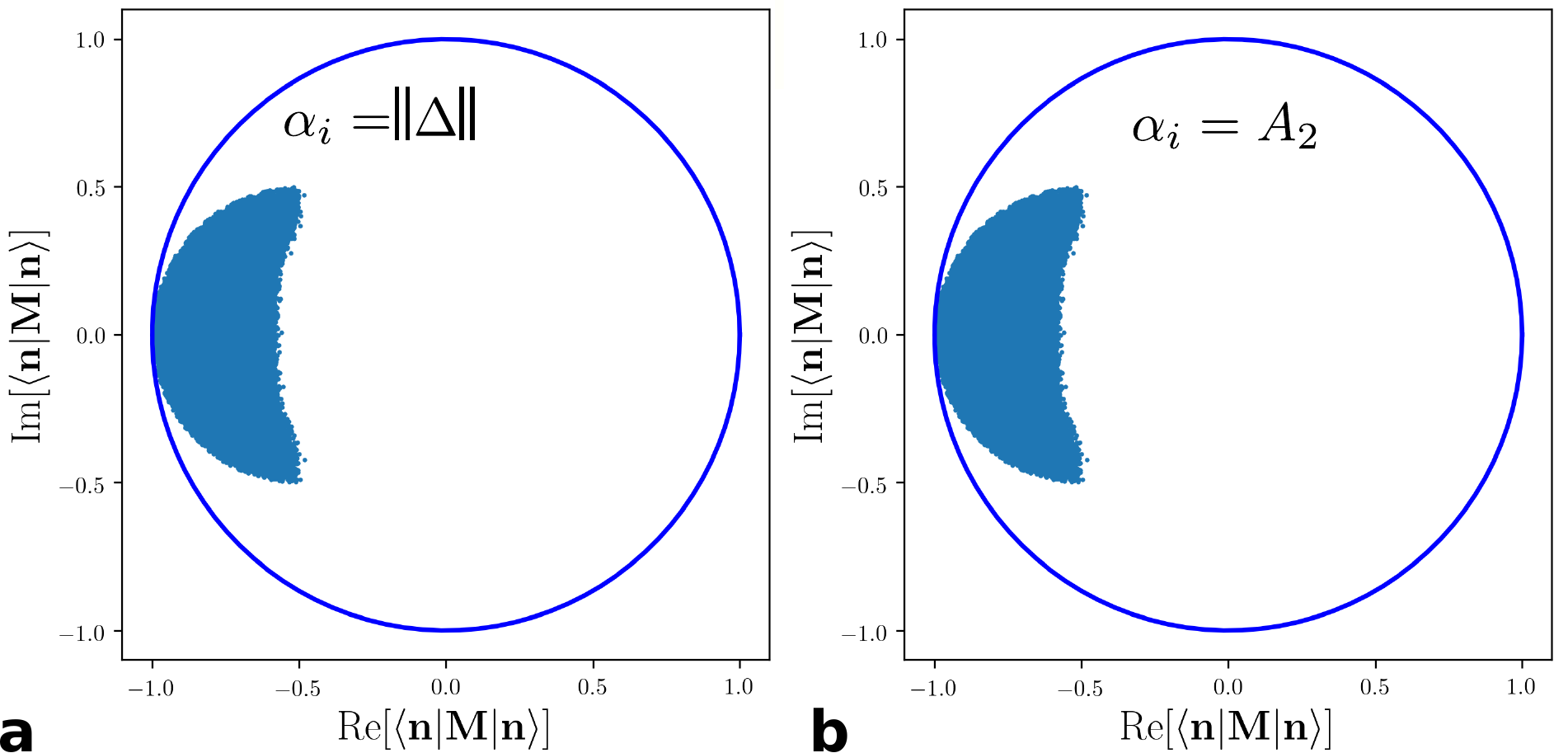}
            }
            \caption{Numerical check that condition $\alpha_\ri > \max\left\{A_1,A_2\right\}$ ensures that the numerical radius of $\boldsymbol{M}$ (given by equation (\ref{eqn:modified_M})) is less than unity. To produce this figure we used the condition on the numerical radius, $\left|\langle \bn|\boldsymbol{M}|\bn\rangle\right|<\langle \bn|\bn\rangle$, with $40\times40$ matrices. We generated $1\times10^6$ random matrices $\bD$ (positive definite $\bD_{\ri}$) and unitary matrices $\boldsymbol{U}$, along with the corresponding values of $\alpha_{\rm i}$ calculated as indicated in each panel (using Python's numpy library for random matrix generation~\cite{Oliphant06}).  For each matrix we calculated values for $|\langle \bn|\boldsymbol{M}|\bn\rangle|$ for a set of 40 random but orthogonal complex vectors, $\bn$. The largest magnitude of these values is one of the $10^6$ points plotted in each panel. In (b) (largest magnitude $0.999997$) we used the condition (\ref{eqn:A2_est}), which we know analytically to be sufficient to move the eigenvalues of $\boldsymbol{M}$ within the unit circle.  In (a) (largest magnitude $0.999999$) we show that $\alpha_{i} > \left\|\Delta\right\|$ appears to guarantee also that the eigenvalues are within the unit circle} 
            \label{fig:numerical_test}
        \end{figure}
      
	      In the main text we showed that the following choice of $\alpha_\ri$ guarantees convergence of the modified Born series
        \begin{equation}
         	\alpha_\ri>\max\left\{A_{1},A_{2}\right\}, \label{eqn:inequality_sup}
        \end{equation}
        where $A_1$ and $A_2$ are defined as:
        \begin{eqnarray}
          A_1 & = & \sqrt{\max_{\bn}\left(\frac{1}{2}\left\langle\bn\right|\bD\mdot\bD^{\dagger}+\bD^{\dagger}\mdot\bD\left|\bn\right\rangle\right)}\nonumber\\
          A_2 & = & \max_{\bn}\frac{|\langle\bn|\bD\mdot\bD_{\ri}+\bD_{\ri}\mdot\bD|\bn\rangle|}{2\langle\bn|\bD_{\ri}|\bn\rangle}
        \end{eqnarray}
        It the main text it was noted that without constraining $\bD_{\ri}$ to be positive definite, $A_{2}$ can be arbitrarily large, when $\langle\bn|\bD_{\ri}|\bn\rangle$ is close to zero.  We begin this Supplementary Section by showing that this divergence can be avoided so long as $\bD_{\ri}$ has an empty kernel (no eigenvectors with zero eigenvalue).  Suppose that $\bD_{\ri}$ has an eigenvector $|\bn_0\rangle$ with eigenvalue $\lambda$, and consider $|\bn\rangle=|\bn_0\rangle+\eta|\bn_\perp\rangle$ where $\eta\ll1$.  The expression for $A_2$ inside the maximization is then given by
%        \begin{multline}
%          \frac{|\langle\bn|\bD\mdot\bD_{\ri}+\bD_{\ri}\mdot\bD|\bn\rangle|}{2\langle\bn|\bD_{\ri}|\bn\rangle} =\\ \frac{|2\lambda\langle\bn_0|\bD|\bn_0\rangle+\eta\lambda(\langle\bn_\perp|\bD|\bn_0\rangle+\langle\bn_0|\bD|\bn_\perp\rangle) +
%            \eta\left(\langle\bn_\perp|\bD_{\ri}\mdot\bD|\bn_0\rangle + \langle\bn_0|\bD\mdot\bD_{\ri}|\bn_\perp\rangle\right)+\eta^{2}\langle\bn_\perp|\bD\mdot\bD_{\ri}+\bD_{\ri}\mdot\bD|\bn_\perp\rangle|}
%            {2\left(\lambda+\eta^{2}\langle\bn_{\perp}|\bD_{\ri}|\bn_{\perp}\rangle\right)}\label{eqn:kernel_of_Di}
%        \end{multline}
        \begin{multline}
        \frac{|\langle\bn|\bD\mdot\bD_{\ri}+\bD_{\ri}\mdot\bD|\bn\rangle|}{2\langle\bn|\bD_{\ri}|\bn\rangle} =\\
        \frac{
        \left|
          \begin{array}{l}
            2\lambda\langle\bn_0|\bD|\bn_0\rangle + \eta\left(\langle\bn_\perp|\bD_{\ri}\mdot\bD|\bn_0\rangle + \langle\bn_0|\bD\mdot\bD_{\ri}|\bn_\perp\rangle\right) + \\
             + \eta\lambda(\langle\bn_\perp|\bD|\bn_0\rangle+\langle\bn_0|\bD|\bn_\perp\rangle) + \eta^{2}\langle\bn_\perp|\bD\mdot\bD_{\ri}+\bD_{\ri}\mdot\bD|\bn_\perp\rangle|
          \end{array}\right|
        }{2\left(\lambda+\eta^{2}\langle\bn_{\perp}|\bD_{\ri}|\bn_{\perp}\rangle\right)}\label{eqn:kernel_of_Di}
        \end{multline}
        If $\lambda=0$ (and the kernel of $\bD_{\ri}$ thus contains $|\bn_0\rangle$ then the above quantity diverges as $1/\eta$ as we take $\eta$ to zero.  The quantity $A_2$ thus becomes infinite. However, if $\lambda$ is non-zero (but arbitrarily small) then as $\eta\to0$, (\ref{eqn:kernel_of_Di}) tends to $|\langle\bn_0|\bD|\bn_0\rangle|$ which is finite, and we note, smaller than or equal to the largest singular value of $\bD$.
        
        Assuming that $\langle\bn|\bD_{\ri}|\bn\rangle$ is never zero, the quantity $A_{2}$ can be rewritten in a form that is easier to compute.  First we write the positive definite Hermitian matrix $\bD_{\ri}$ as the square of another Hermitian matrix $\boldsymbol{a}$
        \begin{equation}
        	\bD_{\ri}=\boldsymbol{a}^{2}
        \end{equation}
        Defining $|\bn'\rangle=\boldsymbol{a}|\bn\rangle$, the quantity $A_{2}$ can be written as
        \begin{equation}
          A_{2}=\max_{\bn'}\frac{|\langle\bn'|\boldsymbol{a}^{-1}\mdot\bD\mdot\boldsymbol{a}+\boldsymbol{a}\mdot\bD\mdot\boldsymbol{a}^{-1}|\bn'\rangle|}{2\langle\bn'|\bn'\rangle}
        \end{equation}
        which is simply the numerical radius $r(\boldsymbol{m})$ of the matrix

        \begin{equation}
        	\boldsymbol{m}=\frac{1}{2}(\boldsymbol{a}^{-1}\mdot\bD\mdot\boldsymbol{a}+\boldsymbol{a}\mdot\bD\mdot\boldsymbol{a}^{-1}).
        \end{equation}
        The numerical radius of a matrix is always less than or equal to its norm~\cite{Dragomir13} $r(\boldsymbol{m})\leq\|\boldsymbol{m}\|$.  In addition, the norm of a sum is never larger than the sum of the norms, and we can therefore estimate $A_{2}$ as
      \begin{equation}
      	A_2=\frac{1}{2}\left[\|\boldsymbol{a}^{-1}\mdot\bD\mdot\boldsymbol{a}\|+\|\boldsymbol{a}\mdot\bD\mdot\boldsymbol{a}^{-1}\|\right]\label{eqn:A2_est}
      \end{equation}
      where $\|\cdot\|$ indicates the $l_2$-norm. While the spectra of $\bD$ and e.g. $\boldsymbol{a}\mdot\bD\mdot\boldsymbol{a}^{-1}$ are the same, their norms are not.  Given that the spectral radius of an operator is always less than or equal to its norm, the lowest possible value of $A_{2}$ is the magnitude of the largest eigenvalue $|\lambda_{\rm max}|$ of $\bD$.  Meanwhile, $A_{1}$, being the square root of the numerical radius of a Hermitian operator, is equal to the square root of the operator norm $(1/2)\|\bD\mdot\bD^{\dagger}+\bD^{\dagger}\mdot\bD\|$, which is bounded by
      \begin{equation}
      	A_{1}\leq\frac{1}{2}\|\bD\mdot\bD^{\dagger}\|+\frac{1}{2}\|\bD^{\dagger}\mdot\bD\|=|\lambda_{\max}|\leq A_{2}
      \end{equation}
      where the estimate for $A_{2}$ is here given by equation~(\ref{eqn:A2_est}). We can therefore use this estimate of $A_{2}$ to find an upper bound for the value of $\alpha_\ri$.  Figure~\ref{fig:numerical_test}b shows a numerical test, where we chose $A_2$ according to (\ref{eqn:A2_est}) and repeatedly evaluated the inner product $\langle\bn|\boldsymbol{M}|\bn\rangle$ for different choices of random complex $40\times40$ matrices $\bD$ and $\boldsymbol{U}$. These tests represent $10^{6}\times40=4\times10^7$ evaluations of the inner product and appear to indicate that $\alpha_\ri > \left\|\bD\right\|$ is a tighter bound (Fig.~\ref{fig:numerical_test}a).   
      
  \section{Impedance matching and Bianisotropy}\label{sec:magnetic_figures}
      %
      %: Impedance matching/chiral figure
      %
      \begin{figure*}[ht!]
        \centering
        \fbox{
          \includegraphics[width=0.97\textwidth]{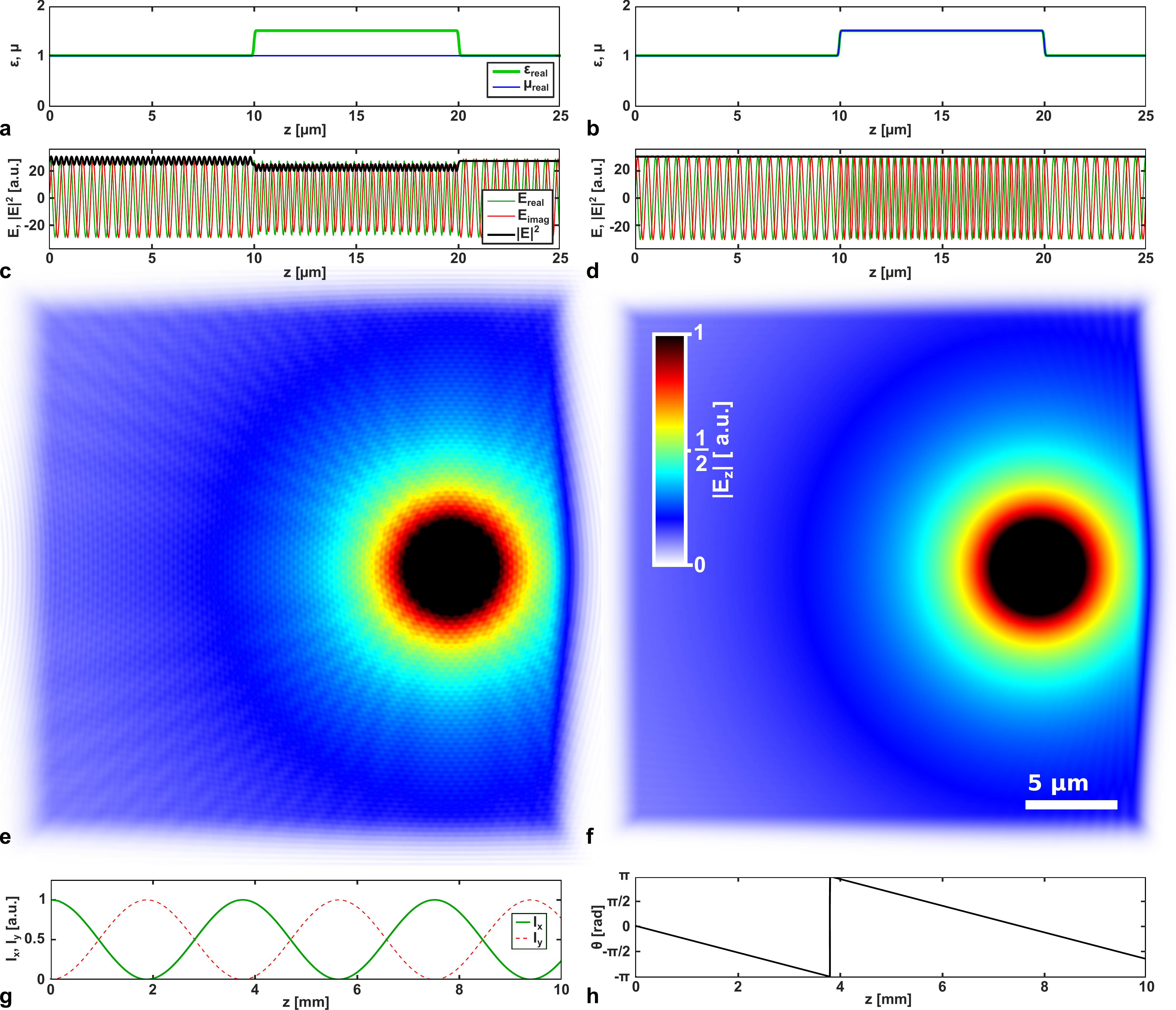}
        }
        \caption{Demonstration of impedance matching (a-f), and propagation in a chiral medium (g,h). A plane wave in free space ($\epsilon=\mu=1$) with a wavelength $\lambda=500\;\mathrm{nm}$ enters at $x=10\mum$ from the left into dielectric slabs of thickness $10\mum$, with $\mu=1$ (a,c) and $\mu=1.5$ (b,d). In both cases the permittivity is $1.5$ (green line, panels a and b). The interference between the incoming and reflected wave is clearly visible as oscillations in intensity ($|E|^2$, black line, c). It can be seen that a fraction of the wave is reflected from the slab without impedance matching ($\mu\not\propto\epsilon$ in panel a). In contrast, a constant intensity is seen in panel (d), indicative of the absence of back-reflection for the impedance matched slab ($\mu \propto \epsilon$ in panel b). To facilitate comparison, both the intensity and field are normalized to their respective maximum value in panels (c) and (d). Panels (e,f) show the (truncated) electric field amplitude for a dipole with absorbing (e) and impedance matched (f) boundary layers. The interference with the back reflected wave, visible as beating in panel (e), is suppressed bt the impedance matched layers as seen in panel (f).
          (g,h) Linear polarization rotates upon propagation in a chiral medium with a high chirality that is 100 times of that of saturated glucose ($n = 1.45$, specific rotation $[\alpha]_{500\;\mathrm{nm}}^T 52.7^\circ\mathrm{mL}\;\mathrm{g}^{-1}\mathrm{dm}^{-1}$, at $909~\mathrm{g/L}$). The constitutive relations can thus be seen to be $\epsilon=\sqrt{1.45}, \mu=1$, and $\xi=\zeta=52.7\frac{909 \lambda}{360}\ri=66.53\times10^{-6}\ri$. The intensity transfer between the x-polarization (solid green) and y-polarization (dashed red) can be seen to occur several times over a propagation distance of $10\;$mm (g). The local angle, $\theta$, of the linear polarization is shown in panel (h). Note the significantly larger length scale for panels (g) and (h).}
        \label{fig:impedanceMatchingAndChiralMedia}
      \end{figure*}
                  
      By eliminating back reflections, impedance matching ensures efficient energy transfer and communication through waveguides. Supplementary Figure~\ref{fig:impedanceMatchingAndChiralMedia}(a-d) introduces the concept of impedance matching in one dimension. Supplementary Figures~\ref{fig:impedanceMatchingAndChiralMedia}(a,b) shows two samples with isotropic permittivity, $\epsilon$, and permeability, $\mu$. Both samples contain objects (at $10\mum < z < 20\mum$) with identical permittivity, larger than the surrounding medium. In the left-hand panel the permeability equals the background, while in the right-hand panel it equals the permittivity for all $z$. Supplementary Figure~\ref{fig:impedanceMatchingAndChiralMedia}(b) thus represents an object that is impedance matched with the surrounding medium. As can been seen from Supplementary Figure~\ref{fig:impedanceMatchingAndChiralMedia}(c), reflections from the front and back surface interfere with the incoming wave. This is most clearly visible in the `beating' in the intensity where the incoming and the reflected wave interfere. Supplementary Figure~\ref{fig:impedanceMatchingAndChiralMedia}(d) shows how impedance matching successfully suppresses back reflections at both interfaces. The absence of oscillations in the field amplitude indicates an absence of back reflections.
      
      Impedance matching also has important practical applications for simulations of infinite volumes in a finite space. 
      Supplementary Figure~\ref{fig:impedanceMatchingAndChiralMedia}(e) shows the electric dipole field in a box with regular absorbing boundaries and homogeneous permeability. As in the one-dimensional case, plotted in Supplementary Figure~\ref{fig:impedanceMatchingAndChiralMedia}(a,c), significant reflections can be noted from the boundaries. Supplementary Figure~\ref{fig:impedanceMatchingAndChiralMedia}(f) shows a dipole in a box with impedance matched absorbing boundaries. It can be seen that the interfering reflections are suppressed. In higher dimensions, impedance matching is only an approximation to perfectly matching layers. This may explain the weak, low-frequency, beating that can be noticed in the top-right corner.
            
      The ability to account for the coupling factors enables the calculation of electromagnetic waves in materials with chiral properties. Supplementary Figures~\ref{fig:impedanceMatchingAndChiralMedia}(g,h) show how a chiral substance slowly rotates the linear polarization of a wave that propagates through it. Supplementary Figure~\ref{fig:impedanceMatchingAndChiralMedia}(g) shows the intensity in the horizontal and vertical polarization component, while Supplementary Figure~\ref{fig:impedanceMatchingAndChiralMedia}(h) shows how the polarization angle changes linearly over the simulation volume's width of $10\;\mathrm{mm}$.
      
  \bibliography{Common}
            
\end{document}

\newcommand{\manuscripttitle}{Calculating coherent light-wave propagation\\ in large heterogeneous media}

\ifthenelse{\boolean{supplement}}{
  \ifthenelse{\boolean{arxiv}}{
  }{
    \setboolean{displaycopyright}{false}
  }
  
%  \title{\manuscripttitle: supplementary material}
%  \title{Calculating coherent light-wave propagation in\\ large heterogeneous media: supplementary material}
  \title{Supplementary Information:\\Calculating coherent light-wave propagation\\ in large heterogeneous media.}
  
  \renewcommand{\thepage}{S\arabic{page}}  
  \renewcommand{\thesection}{S\arabic{section}}
  \renewcommand{\thesubsection}{S\arabic{section}\Alph{subsection}}
  \renewcommand{\thesubsubsection}{S\arabic{section}\Alph{subsection}\Roman{subsubsection}}
  \renewcommand{\thetable}{S\arabic{table}}   
  \renewcommand{\thefigure}{S\arabic{figure}}
  \renewcommand{\thealgorithm}{S\arabic{algorithm}}
  \renewcommand{\theequation}{S\arabic{equation}} 
  
  \renewcommand\refname{Supplementary References}
}{
  \title{\manuscripttitle}
}

\author[1,2,*]{T.~Vettenburg}
\author[1]{S.A.R.~Horsley}
\author[1]{J.~Bertolotti}

\affil[1]{Department of Physics and Astronomy, University of Exeter, Exeter, EX4 4QL, United Kingdom}
\affil[2]{School of Science and Engineering, University of Dundee, Nethergate, Dundee, DD1 4HN, United Kingdom}
\affil[*]{Corresponding author: \href{mailto:t.vettenburg@dundee.ac.uk}{t.vettenburg@dundee.ac.uk}}

\ifthenelse{\boolean{arxiv}}{
  \date{December 25, 2018}
}{
  % To be edited by editor
  \dates{Compiled \today}
  
  %\ociscodes{(050.1755) Computational electromagnetic methods; (080.1753) Computation methods; (080.2720) Mathematical methods (general); (160.1190) Anisotropic optical materials; (260.1440) Birefringence; (160.1585) Chiral media; (160.3918) Metamaterials; (080.2710) Inhomogeneous optical media; (110.7050) Turbid media; (110.0113) Imaging through turbid media; (030.7060) Turbulence; (290.4210) Multiple scattering.}
  \ociscodes{(050.1755) Computational electromagnetic methods; (160.1190) Anisotropic optical materials;\\(260.1440) Birefringence; (160.3918) Metamaterials; (080.2710) Inhomogeneous optical media; (290.4210) Multiple scattering.}
  
  % To be edited by editor
  \newcommand{\thedoi}{10.1364/optica.XX.XXXXXX}
  %\doi{\url{http://dx.doi.org/\thedoi}}
}

%
%\makeatletter
%\def\@maketitle{%
%  \newpage
%  \null
%  \vskip 1em%
%  \begin{center}%
%    \let \footnote \thanks
%    {\HUGE \@title \par}%
%    \vskip 1.0em%
%    {\large
%      \lineskip .5em%
%      \begin{tabular}[t]{c}%
%        \@author
%      \end{tabular}\par}%
%    \vskip 1em%
%    {\large \@date}%
%  \end{center}%
%  \par
%  \vskip 1.0em}
%\makeatother

\makeatletter
\twocolumn[{%
  {\Huge \sffamily \bfseries \fontsize{24}{26} \selectfont \@title \par}%
  \vskip 0.5em %
  {\Huge \sffamily \fontsize{24}{26} \selectfont \@author}%
  \begin{changemargin}{1.5cm}{2.5cm}% 
    \begin{abstract}%
        Understanding the interaction of light with a highly scattering material is essential for optical microscopy of optically thick and heterogeneous biological tissues. Ensemble-averaged analytic solutions cannot provide more than general predictions for relatively simple cases. Yet, biological tissues contain chiral organic molecules and many of the cells' structures are birefringent, a property exploited by polarization microscopy for label-free imaging. Solving Maxwell's equations in such materials is a notoriously hard problem. Here we present an efficient method to determine the propagation of electro-magnetic waves in arbitrary anisotropic materials. We demonstrate how the algorithm enables large scale calculations of the scattered light field in complex birefringent materials, chiral media, and even materials with a negative refractive index.%
    \end{abstract}%
  \end{changemargin}%
  \vskip 1em %
}]
\makeatother

%\maketitle

  \section{Introduction}\label{sec:introduction}
    Determining how light propagates in heterogeneous media is a notoriously hard problem~\cite{Ernst12}. Unless the system of interest has symmetries such as periodicity~\cite{Joannopoulos08}, one needs to solve the Maxwell equations \emph{ab initio} with appropriate boundary conditions. While this may be feasible for relatively simple systems such as Mie scattering~\cite{vandeHulst03}, multiple scattering of light can lead to many subtle effects~\cite{Rotter17,Vellekoop07,Vellekoop10}.
    The fractal propagation method can simulate light in biological tissues~\cite{Glaser16}; however, calculating the exact light field distribution in arbitrary large heterogeneous materials remains out of reach for the current generation of computational methods~\cite{Ernst12}. A further complication is that the media are generally not isotropic, meaning that the refractive index is different depending on the orientation of the field polarization. Such birefringence is common in many samples of interest such as~TiO$_2$, the lipid bilayer cell-membrane~\cite{Salamon01,Everett98,Kuo10}, or muscle fibers~\cite{Jarry00}. Indeed, birefringence is a valuable, label-free, contrast method~\cite{DeBoer97,Jarry00,Alali12,Pierangelo13,Koike-Tani13,Adams16}. Furthermore, biological tissues contain chiral organic molecules such as glucose, a property that is linked to the electro-magnetic coupling and cannot be modeled by anisotropic permittivity. In this paper we investigate the application of the modified Born series~\cite{Osnabrugge16,Krueger17} to solve Maxwell's equations in large heterogeneous electromagnetic media, characterized by arbitrary linear constitutive relations.
    
    Although finite-element and finite-difference time-domain (FDTD) methods can in principle handle general electromagnetic problems of an arbitrary size, such methods do not scale well to the dimensions relevant in microscopy. Only recently it has become possible to represent the complete field distribution in computer memory for larger samples. Conventional methods require amounts of high-speed access storage that are considerably larger. For $\nbpoints$ sample points the finite-element method typically requires the representation of a sparse $3\nbpoints\times 3\nbpoints$ matrix in memory~\cite{Koutromanos18}, which is then iteratively applied in order to reach the desired solution. Meanwhile, the FDTD method requires far higher sampling densities to limit the accumulation of errors. A quite different, and more physically meaningful iterative method is the Born series, which calculates the scattered field, with each iteration including the effect of the next highest order of scattering events. This approach is commonly used for the analytical theory of multiple scattering~\cite{Born26,Akkermans07}. When implemented numerically, all operations can be performed in a constant space ($\propto P$) and using only fast-Fourier transforms and operations on diagonal matrices. The high computational efficiency and modest memory requirements make the Born series an ideal candidate for solving large-scale electromagnetic problems. However, a severe limitation is that the basic form of the series only converges in the limit of weak scattering. A modified Born series for scalar waves, which converges for any value of the scattering strength, was recently proposed by Osnabrugge et~al.~\cite{Osnabrugge16}, and generalized to vector waves by Kr\"uger et al.~\cite{Krueger17}. Yet, as it stands, this approach is limited to media with an isotropic permittivity and without magnetic properties. The modified Born series method thus excludes a large class of materials such as biological tissues that exhibit birefringence or contain chiral organic compounds, as well as any potential applications to metamaterial structures. Here we generalize the numerical Born series method to arbitrary linear materials, including those with heterogeneous magnetic properties and bi-anisotropy. The modification to the Born series is found to be notably more subtle in these materials. We demonstrate that the algorithm enables large scale calculations of the scattered light field in complex birefringent materials, chiral media, and even materials with negative refractive index. Furthermore, we show that the iteration is robust to numerical errors. 

    To start we review the simplest application of the Born series, and its modified form, as given in~\cite{Osnabrugge16,Krueger17}. At a fixed frequency $\omega$, the electric and magnetic fields satisfy Maxwell's equations, $\boldsymbol{\nabla}\times\boldsymbol{E}(\bx) = \ri\omega \boldsymbol{B}(\bx)$, and $\boldsymbol{\nabla}\times\boldsymbol{H}(\bx) = \boldsymbol{j}(\bx) - \ri\omega \boldsymbol{D}(\bx)$, where $\boldsymbol{j}(\bx)$ is the electric current density as a function of the spatial coordinate, $\bx$. In first instance, we consider the case that the medium is isotropic and non-magnetic, and can thus be characterized in terms of a scalar relative permittivity $\epsilon(\bx)$. This quantity connects the displacement field $\boldsymbol{D}$ and magnetic flux density $\boldsymbol{B}$ fields to the electric field $\boldsymbol{E}$ and magnetic field $\boldsymbol{H}$ in the constitutive relations:
  	\begin{align}
  		\boldsymbol{D}(\bx) = \epsilon_0\epsilon(\bx)\boldsymbol{E}(\bx), \;\; \textrm{and} \;\; 
  		\boldsymbol{B}(\bx) = \mu_0\boldsymbol{H}(\bx)\label{eqn:constitutive1} 
  	\end{align}
  	where $\epsilon_0$ and $\mu_0$ are the vacuum permittivity and permeability, respectively. By substituting both constitutive relations into Maxwell's equations we obtain the vector Helmholtz equation for the electric field
    \begin{eqnarray}
      \boldsymbol{\nabla}\times\boldsymbol{\nabla}\times\boldsymbol{E}(\bx) - k_0^2\epsilon(\bx)\boldsymbol{E}(\bx) = \ri\omega\mu_0\boldsymbol{j}(\bx) = \boldsymbol{S}(\bx),\label{eqn:spatially_variant_left_hand_side}
    \end{eqnarray}
    where $k_0=\omega/c$ is the free space wavenumber and $\boldsymbol{S}(\bx)$ represents the radiation source. In order to solve equation~(\ref{eqn:spatially_variant_left_hand_side}) for the electric field $\boldsymbol{E}(\bx)$, one must invert the operator $\boldsymbol{\mathcal{O}}=- k_0^2\epsilon(\bx)+\boldsymbol{\nabla}\times\boldsymbol{\nabla}\times $.
    We could attempt to do this directly, representing $\boldsymbol{\mathcal{O}}$ as a matrix in some basis of eigenfunctions and then applying a numerical inversion algorithm. However, for moderately large systems the direct representation and inversion of such matrix is infeasible, due to demands of both memory and time. Iterative solutions are thus required where, starting from some initial guess for the electric field, one repeatedly applies a matrix until the result is arbitrarily close to the one that would be obtained from a direct inversion, as for instance in finite element simulations (see e.g.~\cite{Koutromanos18}, Chapter 19).

    The Born series~\cite{Newton13} is a physically motivated version of this iterative procedure that was used as a mathematical technique long before the current numerical approaches~\cite{Born26}. It involves splitting the Helmholtz operator, $\boldsymbol{\mathcal{O}}$, into two parts $\boldsymbol{\mathcal{O}}=\Op_h+\Op_i$, where the inverse of $\Op_h$ is known, expanding the rest as what is known as the Born series:
		\begin{multline}
		  \Op^{-1} = \left(\Op_h+\Op_i\right)^{-1} =  \left(\one_3+\Op_h^{-1}\Op_i\right)^{-1}\Op_h^{-1}\\
      =  \left[\one_3-\left(\Op_h^{-1}\Op_i\right)+\left(\Op_h^{-1}\Op_i\right)^{2}+\dots\right]\Op_h^{-1}\\
      = \left[\sum_{\iter=0}^{\infty} \left(-\Op_h^{-1} \Op_i\right)^\iter\right] \Op_h^{-1}, \label{eqn:born_schematic}
		\end{multline}
    where $\one_3$ is the identity operator in three-dimensional space. The operator $\Op_h$ is typically associated with propagation through a homogeneous medium, and $\Op_i$ is due to the inhomogeneity of the material. We can thus understand the Born series (\ref{eqn:born_schematic}) as representing a sequence of scattering events, where as our initial guess the source $\boldsymbol{S}$ generates the same field as if the material were homogeneous $\boldsymbol{E}_0(\bx)=\Op_h^{-1}\boldsymbol{S}$, the first iteration adds to this events where a single scattering event takes place (proportional to $\Op_h^{-1}\Op_i$), the second iteration introduces double scattering (proportional to $(\Op_h^{-1}\Op_i)^2$), and so on. The main problem with this intuitive expansion is that in many cases it will not converge. Physically, this is connected to the existence of bound states~\cite{Newton13}, which are due to the constructive interference of an infinite number of scattering events. On the other hand, the mathematical origin of this divergence is simple, it is the same as the divergence of the scalar series $(1-q)^{-1}=\sum_{\iter=0}^\infty q^\iter, \forall q: q\in\mathbb{C}$: for convergence we must require that $|q|<1$. Correspondingly, the Born series, as given by (\ref{eqn:born_schematic}), will only converge if all of the eigenvalues of $\Op_h^{-1}\Op_i$ have a magnitude less than unity (for discussion see~\cite{Newton13}). Although there are cases where exact results can be deduced from this series~\cite{Horsley15}, the application of the Born series is generally restricted to weakly scattering media.

    For our particular case we take the homogeneous medium of $\Op_h$ to have a constant permittivity $\alpha=\alpha_{\rm r}+{\ri}\alpha_\ri$. The inverse of this operator is given by the Green function for this medium, $\Op_h^{-1}=\bG$, which is the two-index object (dyadic) that is the solution to
		\begin{equation}
      \boldsymbol{\nabla}\times\boldsymbol{\nabla}\times\bG(\bx,\bx')-k_0^{2}\alpha\bG(\bx,\bx')=\one_3\delta^{(3)}(\bx-\bx').\label{eqn:green_eqn}
		\end{equation}
    The remaining part of our operator $\boldsymbol{\mathcal{O}}$~(\ref{eqn:spatially_variant_left_hand_side}), is the spatially-variant difference $\Op_i = -k_0^{2}(\epsilon(\bx)-\alpha)\equiv-k_0^{2}\chi(\bx)$, where $\chi(\bx)$ is the isotropic susceptibility with respect to our choice of background permittivity, $\alpha\in\mathbb{C}$. The Born series, as defined in (\ref{eqn:born_schematic}), for the electric field $\boldsymbol{E}$ then becomes
		\begin{equation}
			\boldsymbol{E} = \Op^{-1} \boldsymbol{S}   %\bG\mdot\boldsymbol{S}+\left(k_0^{2}\bG\mdot\chi\right)\mdot\bG\mdot\boldsymbol{S} + \left(k_0^{2}\bG\mdot\chi\right)^{2}\mdot\bG\mdot\boldsymbol{S}+\dots\\
       = \left[\sum_{\iter=0}^{\infty}(k_0^{2}\bG\mdot\chi)^{\iter}\right]\mdot\bG\mdot\boldsymbol{S}. \label{eqn:modified_born_series}
		\end{equation}
    %where the multiplication represents a sum over both vector and spatial (real or Fourier) indices, and we refer to $\chi$ as an isotropic susceptibility.
    Although we have not yet addressed the problem of convergence, the numerical advantage of this iterative series is that it can be done in a constant space and with a combination of fast-Fourier Transform algorithm~\cite{Walker17}) and operations on matrices that are diagonal in their spatial indices. To see this, we observe that $\chi$ is diagonal in its spatial indices, and that the dyadic Green function (the solution to equation~(\ref{eqn:green_eqn})) can be written as a product of a Fourier transform $\mathcal{F}$, a diagonal matrix (in Fourier space) and an inverse Fourier transform (see supplementary section~S1B)
	\begin{equation}
		\bG=\mathcal{F}^{-1}\mdot\left(\frac{\boldsymbol{\Pi}_{\rm T}}{\boldsymbol{k}^{2}-\alpha k_0^{2}}-\frac{\boldsymbol{\Pi}_{\rm L}}{k_0^{2}\alpha}\right)\mdot\mathcal{F}\label{eqn:dydic_green_function}
	\end{equation}
	where $\boldsymbol{\Pi}_{\rm T}$ and $\boldsymbol{\Pi}_{\rm L}$ are projection matrices that separate the field in a super-position of plane waves with transverse and longitudinal electric-components, respectively. The projection operations are defined as the outer product of the normalized k-vectors, $\boldsymbol{\Pi}_{\rm L}=\boldsymbol{k}\otimes\boldsymbol{k}/\|\boldsymbol{k}\|^2$, and $\boldsymbol{\Pi}_{\rm T}=\one_3-\boldsymbol{\Pi}_{\rm L}$. The quantity in the rounded brackets can thus be seen to be a diagonal matrix in the two Fourier space indices.

  As mentioned above, in a pair of recent papers~\cite{Osnabrugge16,Krueger17}, it was shown that this Born series (\ref{eqn:modified_born_series}) can be made into a convergent series, the only proviso being that the system exhibits solely dissipation (no gain), and that the permittivity is nowhere singular. To achieve this convergence the authors used two ingenious steps. The first is to notice that the complex background permittivity $\alpha$ can be chosen freely, changing the eigenvalues of the operator $k_0^{2}\bG\mdot\chi$, without affecting the solution. The second is that one can apply a pre-conditioner, $\Gamma$, to both sides of the recursion relation~(\ref{eqn:modified_born_series}) as follows:
	\begin{align}
   	%& &\boldsymbol{E} = k_0^2\bG\mdot\chi\mdot\boldsymbol{E}+\bG\mdot\boldsymbol{S}\nonumber\\
    %& \to &
    & \Gamma\boldsymbol{E} = k_0^2\Gamma\bG\mdot\chi\mdot\boldsymbol{E} + \Gamma\bG\mdot\boldsymbol{S}\nonumber\\
     \to & \boldsymbol{E}=(k_0^2\Gamma\mdot\bG\mdot\chi+\one_3-\Gamma)\mdot\boldsymbol{E}+\Gamma\mdot\bG\mdot\boldsymbol{S}
	\end{align}
	The corresponding so-called `modified Born series', which is the counterpart of series~(\ref{eqn:modified_born_series}), is then given by
	\begin{equation}
	  \boldsymbol{E} = \left[\sum_{\iter=0}^{\infty} \boldsymbol{M}^{\iter}\right] \mdot \Gamma\mdot\bG\mdot\boldsymbol{S},~~~\mathrm{where}~~\boldsymbol{M} \equiv k_0^2\Gamma\mdot\bG \mdot\chi-\Gamma+\one_3,\label{eqn:modified_series}
	\end{equation}
	which converges when the absolute value of the largest eigenvalue of $\boldsymbol{M}$ is less than 1. This condition has been shown~\cite{Osnabrugge16,Krueger17} to hold when both the preconditioning operator is given as $\Gamma=\frac{\ri}{\alpha_\ri}\chi$ and when the imaginary part of the background permittivity, $\alpha_\ri$, is larger than the largest value of $|\Delta(\bx)|=\left|\epsilon(\bx) - \alpha_\rr\right|$, considered over all points, $\bx$, in space. The aforementioned work showed that this modified series can be used to very quickly compute the electromagnetic field in large media of moderate index contrast. This result relies on the isotropy of the permittivity, $\epsilon(\bx)$ and on the absence of any magnetic effect or chirality. In this paper we generalize this approach to media with any permittivity or permeability, including anisotropic, magnetic, chiral, and bi-anisotropic media. Interestingly we find that proving the convergence of the modified series seems to be much more subtle when these additional optical effects are accounted for. We also discuss the convergence speed and memory requirements of this approach, and present an easy to implement algorithm to simulate electromagnetic wave propagation in an arbitrary medium. The method is implemented as a \emph{Python 3} library and available with examples for download at \url{https://uod.box.com/v/macromax}.
      
  The paper is organized as follows. In section~\ref{sec:anisotropic_dielectric} we describe how the modified Born series method can be generalized to anisotropic dielectrics. The iterative algorithm is introduced and we discuss how the preconditioner values can be calculated to ensure convergence. The algorithm is demonstrated with the polarization-dependent propagation through homogeneous and highly heterogeneous birefringent materials. Next, we show that the method is not limited to Hermitian permittivity matrix-functions and that it can be extended to non-Hermitian permittivity. In Section~\ref{sec:magnetic_and_chiral_materials} we show how the preconditioning step of the algorithm can be adapted so that the same iteration can be used for magnetic materials. Accounting for electric-magnetic coupling also enables chiral, Tellegen~\cite{Tellegen48}, and general bi-anisotropic media. This is demonstrated with the propagation of a linearly polarized wave of visible light through $10\;$mm of a highly chiral substance. Finally we show how the method naturally handles more esoteric materials such as a solid with a negative refractive index, thus demonstrating the capability of simulating light propagation in such metamaterials.
 
  \section{A convergent Born series for anisotropic dielectrics}\label{sec:anisotropic_dielectric}  
  	As a first step towards a Born series for arbitrary electromagnetic materials, we consider non-magnetic birefringent materials. Optical elements such as waveplates and Wollaston prisms are birefringent and characterized by anisotropic permittivity that cannot be represented by a scalar function, $\epsilon(\bx)$. In this section we show how the modified Born series can be extended to permittivity tensors. Later it will be shown that arbitrary electromagnetic problems can be brought into the same form and solved using the iterative method introduced in this section.
  
    \begin{figure*}[ht!]
      \centering
        \vbox{
        \includegraphics[width=0.97\textwidth]{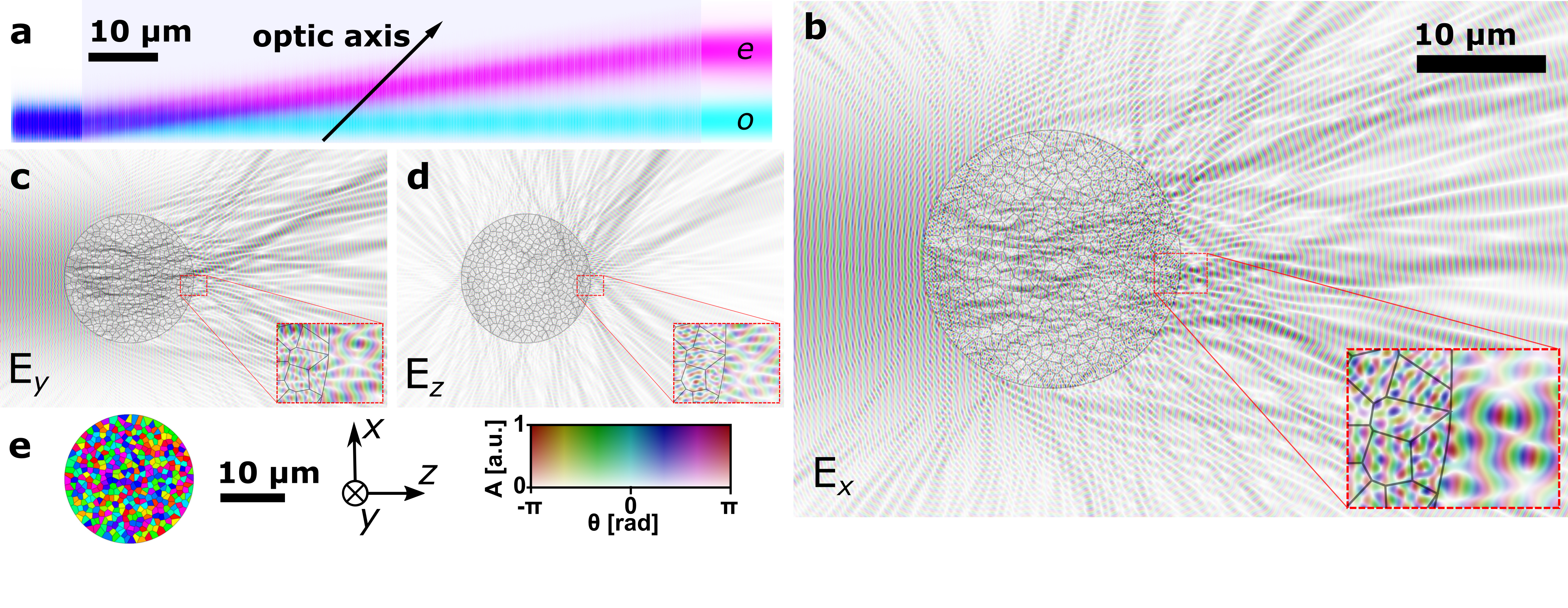}
      }
      \caption{(a) Demonstration of anisotropic permittivity. Diagonally polarized light propagates from left to right through a calcite crystal (light gray box) cut at $45^\circ$ with respect to its optic axis (indicated by the arrow).
      It can be seen that, as expected, the in-plane polarized extraordinary ray (e, magenta) is displaced from the ray that is polarized perpendicular to the plane (o, cyan). Some interference can be noticed between the incoming wave and its back-reflection at both the entrance and exit surface of the crystal.
      (b-e) A circularly polarized Gaussian beam incident from the left on a birefringent vaterite (CaCO$_3$) microrod with a diameter of $20\mum$ forms a complex scattering pattern instead of a single focus. Although the volume is homogeneous CaCO$_3$, complex, seemingly random, scattering occurs due to its subdivision in crystals of approximately $1\mum$ in cross-section for which the fast axis is oriented randomly with angles $\theta$ shown as hue in panel (e). Panels (b-d) show the field components $E_x$, $E_y$, and $E_z$, respectively. The darkness and hue indicate the field amplitude and phase, respectively, as indicated by the legend in panel (e). An overlaid gray grid outlines the crystal areas for reference, and the inset shows a $4\times$ magnified detail of the field at the exit surface.}
      \label{fig:birefringence}
    \end{figure*}
  
    To account for anisotropy, the scalar permittivity $\epsilon(\bx)$ is replaced by the $3\times 3$ matrix $\boldsymbol{\epsilon}(\bx)$ function in the spatial coordinates, $\bx$. In this more general case the constitutive relation for the electric displacement field (\ref{eqn:constitutive1}) becomes
    \begin{eqnarray}
      \boldsymbol{D}(\bx) & = & \epsilon_0\boldsymbol{\epsilon}(\bx)\mdot\boldsymbol{E}(\bx),
    \end{eqnarray}
    and similarly the susceptibility becomes $\boldsymbol{\chi}(\bx)=\boldsymbol{\epsilon}(\bx) - \alpha\one_3$. Applying this constitutive relation to the Maxwell equations as we did to obtain (\ref{eqn:spatially_variant_left_hand_side}), the vector Helmholtz equation becomes
    \begin{eqnarray}
    	\boldsymbol{\nabla}\times\boldsymbol{\nabla}\times\boldsymbol{E}-k_0^2\alpha\boldsymbol{E}(\bx)-k_0^2\boldsymbol{\chi}(\bx)\mdot\boldsymbol{E}(\bx) & = & \boldsymbol{S}(\bx).\label{eqn:vector_helmholtz}
    \end{eqnarray}    
    An anisotropic analogue of the modified Born series is given by replacing the scalar preconditioner with a matrix function, $\boldsymbol{\Gamma}(\bx)\equiv\frac{\ri}{\alpha_\ri}\boldsymbol{\chi}(\bx)$, in equation~(\ref{eqn:modified_series}). The iteration matrix can now be written as a function of the matrix function $\boldsymbol{\chi}(\bx)$:
    \begin{eqnarray}
    	\boldsymbol{M} & \equiv & \frac{{\ri}k_0^{2}}{\alpha_\ri}\boldsymbol{\chi}\mdot\bG\mdot\boldsymbol{\chi}-\frac{\ri}{\alpha_\ri}\boldsymbol{\chi}+\one_3.\label{eqn:M_operator}
    \end{eqnarray}
    
    The question is whether there still exists a choice of $\alpha_\ri$ such that all the eigenvalues of $\boldsymbol{M}$ have a magnitude less than unity. To show that there is such a choice of $\alpha_i$ we consider the numerical radius, $\max_{\bn}\left|\langle \bn|\boldsymbol{M}|\bn\rangle\right|$, where $\langle\bn|\bn\rangle=1$. Evidently the numerical radius will always be at least as large as the largest magnitude eigenvalue of $\boldsymbol{M}$ and thus we can enforce the convergence of the series (\ref{eqn:modified_series}) through the requirement
    \begin{eqnarray}
    	\max_{\bn}\left|\left\langle \bn\left|\frac{{\ri}k_0^{2}}{\alpha_\ri}\boldsymbol{\chi}\mdot\bG\mdot\boldsymbol{\chi}-\frac{\ri}{\alpha_\ri}\boldsymbol{\chi}+\one_3\right|\bn\right\rangle\right| & < & 1\label{eqn:numerical_radius_condition}
    \end{eqnarray}
    In Supplementary Section~S2 it is shown that this requirement can be satisfied by choosing $\alpha_\ri$ to be larger than $\|\bD\|$, the largest singular value of $\bD \equiv \boldsymbol{\epsilon} - \alpha_\rr\one_3$, under the conditions that the eigenvectors of $\bD$ are orthogonal and that the material is free of gain with non-zero losses. Orthogonal eigenvectors are common in practice, e.g.~when there is no point with both anisotropic absorption and birefringence in the material. In what follows we further show that a sufficiently large value of $\alpha_\ri$ can ensure convergence for any material with non-zero losses.
        
    The dyadic Green function, $\bG$, in convergence condition~(\ref{eqn:numerical_radius_condition}) can be written in terms of the identity and an unitary operator $\boldsymbol{U}$ as $\bG \equiv -(2{\ri}k_0^2\alpha_i)^{-1}(\one_3-\boldsymbol{U})$ (see Supplementary Section~S1B). Condition~(\ref{eqn:numerical_radius_condition}) can then be rewritten using the triangle inequality $|a+b|\leq |a|+|b|$ as
   \begin{eqnarray}
    	\max_{\bn}\left[\left|\langle \bn|\bD^{2}-\alpha_\ri^{2}\one_3|\bn\rangle\right|+\left|\left\langle \bn\left|\boldsymbol{\chi}\mdot\boldsymbol{U}\mdot\boldsymbol{\chi}\right|\bn\right\rangle\right|\right] & < & 2\alpha_\ri^{2} \label{eqn:numerical_radius_condition_without_G}
    \end{eqnarray}
    where $\bD=\boldsymbol{\chi}+{\ri}\alpha_\ri\one_3$. To simplify this condition further we apply the Cauchy-Schwarz inequality $|\langle\boldsymbol{a}|\boldsymbol{b}\rangle|^{2}\leq\langle \boldsymbol{a}|\boldsymbol{a}\rangle\langle\boldsymbol{b}|\boldsymbol{b}\rangle$ to remove the unitary operator from the second term, followed by the inequality $a^2+b^2\geq 2 ab$
    \begin{multline}
    	\left|\langle\bn|\bD^{2}-\alpha_\ri^{2}|\bn\rangle\right|+\left\langle\bn\left|\frac{1}{2}(\bD\mdot\bD^{\dagger}+\bD^{\dagger}\mdot\bD)\right|\bn\right\rangle-2\alpha_\ri\langle\bn|\bD_\ri|\bn\rangle\\
       < \alpha_\ri^{2}
      \;\;\;\;\forall \bn: \|\bn\|=1,
    \end{multline}
    where $\bD=\bD_\rr+{\ri}\bD_\ri$, with $\bD_\rr$ and $\bD_\ri$ Hermitian matrices which depend respectively on the reactive and dissipative response of the material. As a final step we rewrite $\bD^{2}=(1/2)(\bD\mdot\bD^{\dagger}+\bD^{\dagger}\mdot\bD)+{\ri}(\bD\mdot\bD_\ri+\bD_\ri\mdot\bD)$ and apply the triangle inequality for a second time
    \begin{multline}
    	\left|\left\langle \bn\left|\frac{1}{2}\left(\bD\mdot\bD^{\dagger}+\bD^{\dagger}\mdot\bD\right)\right|\bn\right\rangle-\alpha_\ri^{2}\right|+\left|\left\langle \bn\left|\bD\mdot\bD_\ri+\bD_\ri\mdot\bD\right|\bn\right\rangle\right|\\
      +\left\langle \bn\left|\frac{1}{2}\left(\bD\mdot\bD^{\dagger}+\bD^{\dagger}\mdot\bD\right)\right|\bn\right\rangle-2\alpha_\ri\langle \bn|\bD_\ri|\bn\rangle < \alpha_\ri^2\\
    \;\;\;\; \forall \bn: \|\bn\|=1.\label{eqn:final_inequality}
    \end{multline}
    We are now in a position to give a condition for the convergence of our modified Born series. Assuming that $\bD_\ri$ is a positive definite operator everywhere, equivalent to assuming a gain-free material that exhibits a dissipative response throughout space, we can restrict $\alpha_\ri$ to be greater than the larger of the following two quantities:
    \begin{align}
    A_1 = & \max_{\bn}\sqrt{\frac{1}{2}\left\langle \bn\left|\bD\mdot\bD^{\dagger}+\bD^{\dagger}\mdot\bD\right|\bn\right\rangle},\\
    A_2 = & \max_{\bn}\frac{|\langle \bn|\bD\mdot\bD_\ri+\bD_\ri\mdot\bD|\bn\rangle|}{2\langle \bn|\bD_\ri|\bn\rangle}.
    \end{align}
    If the convergence condition
    \begin{eqnarray}
     \alpha_\ri & > & \max\left\{A_1,A_2\right\},\label{eqn:convergence_condition_general}
    \end{eqnarray}
    is fulfilled then the inequality (\ref{eqn:final_inequality}) will be satisfied, and the modified Born series will converge for the anisotropic medium in question.
    
    For this particular bound it is very important that $\bD_\ri$ is positive definite, rather than non-negative. If $\bD_\ri$ has a kernel containing even one eigenvector $|\bn_0\rangle$ then for vectors $|\bn\rangle=|\bn_0\rangle+\eta|\bn_{\perp}\rangle$, $A_2$ diverges as $\sim 1/\eta$, as $\eta\to0$. This is because $A_2$ is the analogue of a weak value~\cite{Aharonov88} of $\bD_{\rr,\ri}$ with respect to the two vectors $|\bn\rangle$ and $\bD_\ri|\bn\rangle$, a value which is well known to potentially lie outside the spectrum of the operator when the two vectors are close to orthogonal (so-called superweak values~\cite{Berry10}).
    In Supplementary Section~S2B it is shown that when $\bD_\ri$ has an empty kernel, no such divergence arises, and the Born series can be made to converge for any anisotropic medium. As it stands, it is difficult to make use of condition~(\ref{eqn:convergence_condition_general}), because the value of $A_2$ cannot be readily estimated. Extensive numerical tests (Supp.~Sec.~S2B) have also indicate the larges singular value $\left\|\bD\right\|$ is a tighter bound for $\alpha_\ri$ that is valid for general gain-free materials, although we have not been able to show this analytically. As a diverging series is straightforward to detect, we have set $\alpha_\ri = \left\|\bD\right\|$ in all our simulations. Note that in the limiting case where the medium is isotropic and non-magnetic, $\bD=\Delta\one_3$, and the basis $|\bn\rangle$ is the position basis, convergence condition~(\ref{eqn:convergence_condition_general}) reduces to $\alpha_\ri > \max_{\boldsymbol{x}}|\Delta(\boldsymbol{x})|$. This reproduces the convergence condition found by both Osnabrugge et al.~\cite{Osnabrugge16} and Kr\"uger et al.~\cite{Krueger17}. %Numerical checks of our generalized convergence criterion~(\ref{eqn:convergence_condition_general}) are provided in Supplementary Section~S2B.

    %\subsection{The general iteration}\label{sec:general_iteration}
      To demonstrate that this series can be used to calculate the propagation of light through an anisotropic material, we consider a birefringent crystal. Perhaps the most common example is calcite (CaCO$_3$, $n_o=2.776$, $n_e=2.219$ at $\lambda=500\;\mathrm{nm}$), which splits an incident beam into two orthogonally-polarized beams that travel along different paths~\cite{Waldman02}. Fig.~\ref{fig:birefringence}a shows that the modified Born series reproduces this effect for a circularly-polarized Gaussian beam (wavelength of $500\;\mathrm{nm}$) traversing the crystal from left to right. The crystal displaces the extra-ordinary polarization laterally ($e$, magenta), while the ordinary-polarized component ($o$, cyan) travels along the original optical axis unaffected by the anisotropy.
      
      \begin{figure*}[ht!]
        \centering
          \fbox{
            \includegraphics[width=0.97\textwidth]{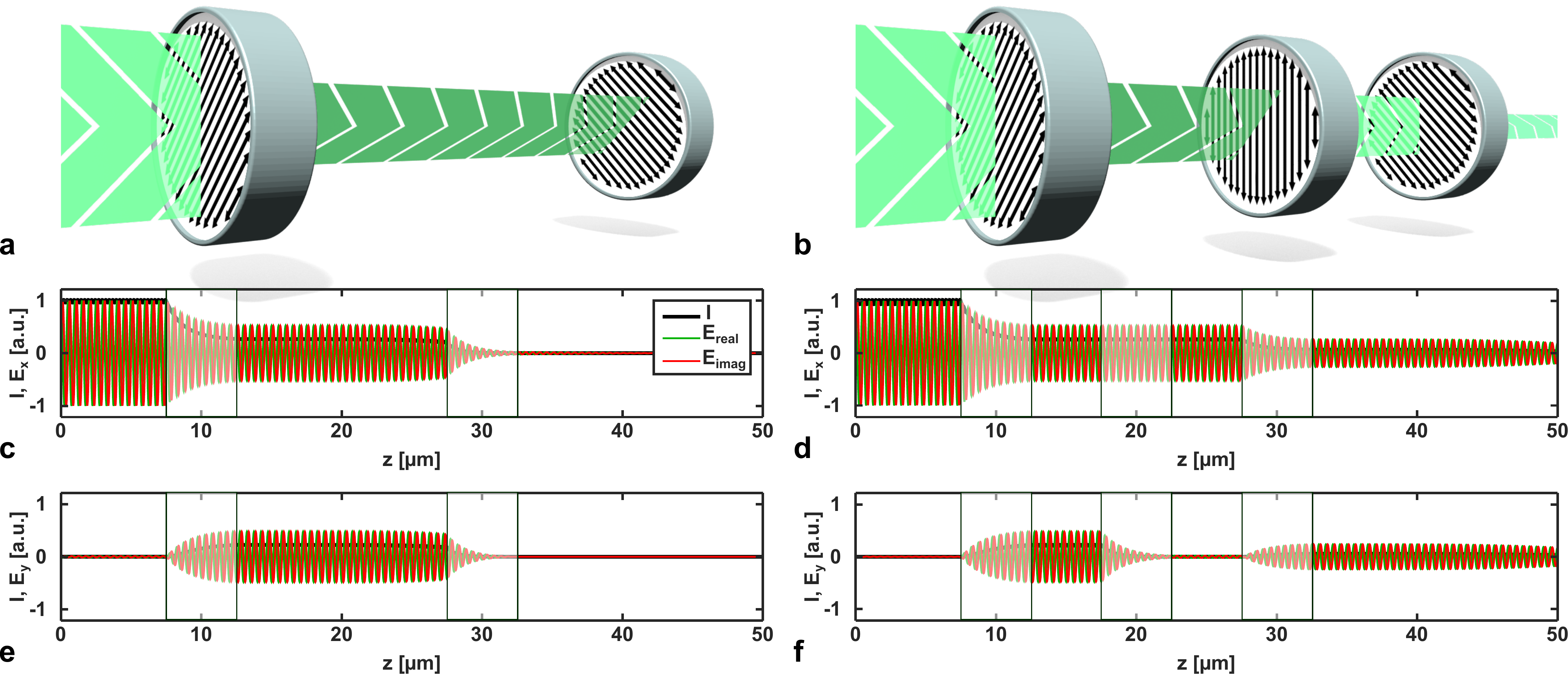}
          }
          \caption{Demonstration of non-hermitian anisotropy. A x-polarized wave traverses two (a,c,e), or three (b,d,f), polarizers from the left to the right. The first and the last polarizer are in cross-diagonal-orientation, preventing transmission through the first system shown in (a). The electric field and intensity are shown for the vertical and horizontal components in (c,e) and (d,f), respectively. Arbitrary units are used so that the maximum value of the intensity and field match for clarity of display.}
          \label{fig:crossPolarizers}
      \end{figure*}
      
    	This method of computing the solution to Maxwell's equation is quite unusual compared to, for instance, a finite difference calculation. Firstly, each iteration can be performed in constant space and involves only a product of diagonal matrices and Fourier transforms, both of which have relatively low computational demands. Secondly, and perhaps even more interesting, the iterative corrections to the field predominantly depend on the local field. This is because the modified Green function, $\bG(\bx,\bx')$, is the solution to equation~(\ref{eqn:green_eqn}), which is that for a dissipative medium characterized by a positive $\alpha_{\ri}$. Both of these features can be advantageous for the efficient simulation of wave propagation in large heterogeneous materials. To illustrate this we simulated ($\lambda=500\;\textup{nm}$) the electric field within and behind a heterogeneous calcite rod of diameter $20\mum$ made up of crystal grains with a variable diameter of approximately $1\mum$ and with a random orientation (Fig.~\ref{fig:birefringence}b-e). An isotropic material would have focused both polarizations. Instead, a highly irregular speckle pattern can be seen to emerge, elongated along its propagation direction. The insets show close-ups of the high irregularity in the three field-components as the beam exits at the CaCO$_3$-air interface.

      In the case of ordinary birefringence, the $3\times 3$-matrix that represents the permittivity at a specific point in space is Hermitian. More general materials may have a permittivity that is non-Hermitian. To illustrate the application of the modified Born series to lossy anisotropic media we calculated the transmission through the simplest such system: an absorbing polarizer.
      Fig.~\ref{fig:crossPolarizers} shows a series of polarizers with varying alignment and an anisotropic extinction coefficient of $\kappa=0.1$. For clarity, back reflections are avoided by setting the refractive index of the polarizers to the same value as that of the embedding medium ($n=1$). 
      The two geometries shown are two cross-polarizers (a,c,e), the second of which blocks all transmission. When a third polarizer is inserted in-between the same two polarizers, and with its transmission axis at $45^\circ$ to the first two polarization axes (b,d,f), this results in a non-zero transmission in agreement with Malus' law.
      
      Algorithm~\ref{alg:electric} shows the pseudo code to calculate the electric field in anisotropic non-magnetic materials. Before starting the iteration loop, the algorithm must choose a background permittivity, $\alpha$ that will lead to a fast convergence. This is done by numerically finding the value $\alpha_\rr$ that minimizes the largest singular value of $\bD=\boldsymbol{\epsilon}(\bx) - \alpha_\rr\one_3$, using the Nelder-Mead algorithm~\cite{Nelder65}. The imaginary part of the background permittivity, $\alpha_\ri$, is then set to equal the minimized upper bound for the singular value $\|\bD\|$. Although the strict equality does not necessarily guarantee convergence, in practice we have never encountered divergence. It should be noted that calculating the largest singular value is impractical for large problems, so an upper bound must be calculated instead and $\alpha_\ri$ will likely be larger than the largest singular value, $\|\bD\|$. For good measure, the complete algorithm~(S1) presented in Supplementary Section~S1, does include a check for divergence, though in practice it has never been encountered. The algorithms' description is followed by a discussion of its efficiency, robustness to errors, and the issues of sampling and aliasing.
      
      The iteration loop starts on line~\ref{op:electric_iteration_begin} and repeats until the condition on line~\ref{op:electric_iteration_end} is met. In each loop a correction term, $\Delta\boldsymbol{E}$, is calculated for the electric field estimate, $\boldsymbol{E}$. Unless divergence is detected, the correction term is added to the current electric field estimate. In the unlikely event that divergence is detected, the current iteration is repeated for a more conservative, larger, imaginary value for the permittivity bias, $\alpha_\ri$.
      This procedure is repeated until the correction term becomes smaller than a preset tolerance, $r_{\max}$. Alternative stopping criteria may be considered to improve the algorithm's performance~\cite{Krueger17}.
      
      \begin{algorithm}
       \caption{Calculation of the electric field in materials with anisotropic permittivity, $\boldsymbol{\epsilon}$.}\label{alg:electric}
        \begin{algorithmic}[1]
          \Function{SolveAnisotropic}{$\Delta\boldsymbol{r}, \boldsymbol{j}, \boldsymbol{\epsilon}, \boldsymbol{E_0}=\boldsymbol{0},r_\mathrm{max}$}
          \State $\alpha_\rr \gets \argmin{\alpha_\rr}\left[\left\|\boldsymbol{\epsilon} - \one_3\alpha_\rr\right\|\right]\;\;\;\;\forall \alpha_\rr \in \mathbb{R}$
          \State $\alpha_\ri \gets \left\|\boldsymbol{\epsilon} - \one_3\alpha_\rr\right\|$
          \State $\boldsymbol{\chi}  \gets  \boldsymbol{\epsilon} - \left(\alpha_\rr + \ri\alpha_\ri\right)\one_3$
          \State $\boldsymbol{E} \gets \boldsymbol{E_0}$
          %\State $p \gets \infty$ \Comment{The algorithm keeps track of the l$_2$-norm of the previous update to detect divergence.}
          \Repeat \label{op:electric_iteration_begin}
            \State $\Delta\boldsymbol{E} \gets \frac{\ri}{\alpha_\ri}\boldsymbol{\chi} \mdot\left[\bG\ast \left(k_0^2\boldsymbol{\chi}\mdot\boldsymbol{E} + \boldsymbol{S}\right) - \boldsymbol{E}\right] $ \label{op:calc_update_electric}
              \State $\boldsymbol{E} \gets \boldsymbol{E} + \Delta\boldsymbol{E}$
          \Until{$\left\|\Delta\boldsymbol{E}\right\| < r_\mathrm{max} \left\|\boldsymbol{E}\right\|$} \label{op:electric_iteration_end}
          \State \Return $\boldsymbol{E}$
          \EndFunction
        \end{algorithmic}
      \end{algorithm}

  %
  %: General electromagnetic media
  %
  \section{Magnetic and chiral materials}\label{sec:magnetic_and_chiral_materials}
    As it stands, Algorithm~\ref{alg:electric} can only be used on non-magnetic materials. However, the preconditioning step can be generalized to account for general magnetic media, without requiring changes to the iteration loop. A general algorithm is given in Supplementary Section~S1A. The susceptibility, $\boldsymbol{\chi}$, must be redefined for magnetic materials to account for permeability and any electro-magnetic coupling constants. A permeability scale, $\beta$, is introduced, and the background permittivity, $\alpha$, is determined from the modified susceptibility. 
    
    %\subsection{Preconditioning for general magnetic materials}
    The constitutive relations for general linear electromagnetic materials are given by (see e.g.~\cite{Priou12}):
    \begin{eqnarray}
      \boldsymbol{D}(\bx) &=& \epsilon_0\boldsymbol{\epsilon}(\bx)\mdot\boldsymbol{E}(\bx)+\frac{1}{c}\boldsymbol{\xi}(\bx)\mdot\boldsymbol{H}(\bx)\nonumber\\
      \boldsymbol{B}(\bx) &=& \mu_0\boldsymbol{\mu}(\bx)\mdot\boldsymbol{H}(\bx)+\frac{1}{c}\boldsymbol{\zeta}(\bx)\mdot\boldsymbol{E}(\bx).\label{eqn:general_constitutive}
    \end{eqnarray}
    Here both the permittivity $\boldsymbol{\epsilon}(\bx)$ and permeability, $\boldsymbol{\mu}(\bx)$, are tensors that depend on the position, $\bx$. The additional coupling terms $\boldsymbol{\xi}(\bx)$ and $\boldsymbol{\zeta}(\bx)$ enable arbitrary linear interactions between the electric and magnetic components, which are commonplace in---for instance---metamaterials~\cite{Smith06}. The coupling tensors $\boldsymbol{\xi}(\bx)$ and $\boldsymbol{\zeta}(\bx)$ are also essential to model chiral materials~\cite{Lakhtakia94}, and non-reciprocal materials such as moving media~\cite{Landau04}, and Tellegen media~\cite{Tellegen48}.
    
    Substituting (\ref{eqn:general_constitutive}) into the Maxwell equations and eliminating the magnetic field, $\boldsymbol{H}=(\ri\omega\mu_0)^{-1}\boldsymbol{\mu}^{-1}\mdot[\boldsymbol{\nabla}\times\boldsymbol{E}-\ri k_0\boldsymbol{\zeta}\mdot\boldsymbol{E}]$, leads to the same form of vector Helmholtz equation~(\ref{eqn:vector_helmholtz}). It is sufficient to generalize the source and susceptibility as follows:
    \begin{eqnarray}
      \boldsymbol{S} & \equiv & \ri\omega\mu_0\beta^{-1}\boldsymbol{j}\label{eqn:general_source}\\
      \boldsymbol{\chi} & = & \frac{\boldsymbol{\epsilon}}{\beta}-\frac{1}{\beta}\boldsymbol{\xi}\mdot\boldsymbol{\mu}^{-1}\mdot\boldsymbol{\zeta} - \one_3\alpha + -\frac{\ri}{\beta}\boldsymbol{\xi}\mdot\boldsymbol{\mu}^{-1}\mdot\CurlOp \nonumber\\
      & &  + \frac{\ri}{\beta}\CurlOp\mdot\boldsymbol{\mu}^{-1}\mdot\boldsymbol{\zeta} + \CurlOp\mdot\left(\one_3 - \frac{\boldsymbol{\mu}^{-1}}{\beta}\right)\mdot\CurlOp,\label{eqn:general_susceptibility}
    \end{eqnarray}
    where $\CurlOp\mdot \equiv k_0^{-1}\boldsymbol{\nabla}\times$ is the differential operator, which calculation is discussed in Supplementary Section~S1A. For conciseness, the spatial dependency of the variables is omitted in equations~(\ref{eqn:general_source}) and~(\ref{eqn:general_susceptibility}).
    Note that all these operations can be implemented efficiently using fast-Fourier transforms and diagonal-matrix multiplications.
    Since vector Helmholtz equation~(\ref{eqn:vector_helmholtz}) still applies, the same iterative procedure can be used to solve for non-magnetic anisotropic materials.

  	Although the modified Born series is now formally identical for these general constitutive relations (\ref{eqn:general_constitutive}), to that for the anisotropic dielectric discussed in section~\ref{sec:anisotropic_dielectric}, it is important to note that the susceptibility (\ref{eqn:general_susceptibility}) is no longer a diagonal matrix in the position variable. This is because it now contains the differential operators $\CurlOp$. Although this affects neither the form of the modified Born series (i.e. as a product of diagonal matrices and Fourier transforms), or the proof of its convergence (which did not make any assumptions about the form of $\boldsymbol{\chi}$), it can make the convergence of the series less rapid. This is because the largest singular value of $\CurlOp$ has a magnitude of $\pi/(k_0\Delta r)$, where $\Delta r$ is the grid spacing. Therefore, the required number of iterations increases with the density of the sampling grid when simulating general magnetic materials.

    The optimal value of $\alpha$ is determined using a similar method in both the all-dielectric and general cases. Albeit not critical for convergence, the additional parameter $\beta$ we have introduced in (\ref{eqn:general_susceptibility}) can be used to enable further optimization of the convergence speed.
   
    Algorithm~S1 in Supplementary Section~S1A generalizes Algorithm~\ref{alg:electric} to magnetic materials by redefining $\hat{\sigma}$ as a function of two variables, $\alpha_\rr$ and $\beta$, and minimizing it in both variables.
    Using the usual inequalities for matrix norms~\cite{Gradshteyn00}, the function $\hat{\sigma}(\beta,\alpha_\rr)$ determines an upper bound for the singular values of the separate terms in the sum (\ref{eqn:general_susceptibility}), and using the largest singular value, $\sigmaCurl = \pi/(k_0\Delta r)$, of the discretized differential operator, $\CurlOp$.
    The Nelder-Mead method is used to numerically minimize the product $\hat{\sigma}(\beta, \alpha)\beta$, which we found to correlate well with the inverse of the convergence rate. The minimization provides values for the permeability scale, $\beta$, and the real part of the background permittivity $\alpha_\rr$. The imaginary part, $\alpha_\ri$, is set to $\hat{\sigma}(\beta, \alpha)$, an upper bound estimate for the largest singular value of $\bD = \boldsymbol{\chi} + \ri\alpha_\ri\one_3$.
      
    %subsubsection{Impedance Matched Layers}
    At optical frequencies, it is a common approximation to assume that the permeability, $\boldsymbol{\mu}$, equals $\mu_0\one_3$ in the entire simulation volume. While this is appropriate at such frequencies (where magnetism can in many circumstances be ignored~\cite{Landau04,Merlin09}), the modified Born series can---in principle---be applied to any electromagnetic problem, where magnetic effects are important. Indeed, while all our simulations are apparently for optical length scales we must remember that Maxwell's equations are scale invariant so that the same results would be obtained with the spatial frequencies, $\bk$, scaled by the same factor as the temporal frequency, $\omega$. In Supplementary Figure~S4(a-f), we show that our generalized Born series can simulate impedance matched media (where $\boldsymbol{\epsilon}=\boldsymbol{\mu}$, which includes transformation media~\cite{Pendry06}, as well as other interesting wave effects (e.g.~\cite{Vial16}).
    
    Since the constitutive relations (\ref{eqn:general_constitutive}) include the electric-magnetic coupling tensors, $\boldsymbol{\zeta}$ and $\boldsymbol{\xi}$, one can also use the modified Born series to treat bi-isotropic and more generally bi-anisotropic media. Many organic molecules such as glucose have a chiral asymmetry that leads to bi-isotropy. This chirality causes a rotation of linear polarization around the axis of propagation. As a second example of our generalized approach, Supplementary Figures~S4(g and h) show simulations where linearly polarized light is slowly rotated in a chiral solution. As the rotation takes several millimeters to complete, the calculation was performed for a wavelength of $500\;\mathrm{nm}$ and over a propagation length of $10\;\mathrm{mm}$.
    
    %\subsubsection{Chiral media as an example of bi-isotropy}
  		
\begin{figure}[htbp]
    	\centering
    	\fbox{
    		\includegraphics[width=0.97\linewidth]{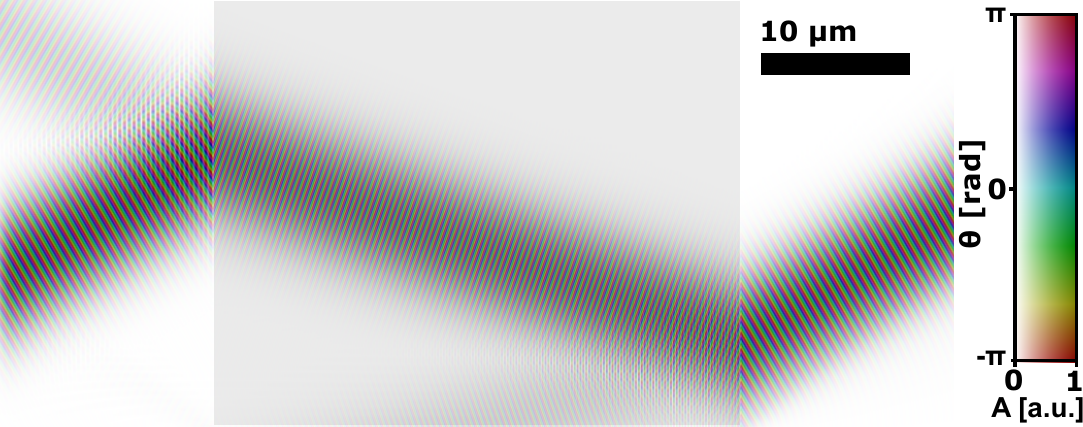}
    	}
    	\caption{Light-wave propagation through a material with negative refractive index ($\epsilon = -1.5$, $\mu = -1$, gray). The field amplitude (brightness) and phase (hue) are shown for a Gaussian beam that enters the surface at $30^\circ$ to the normal. The beam can be seen to refract at $-19.5^\circ$, backwards, into the metamaterial at an angle opposite to that for normal glass.}
    	\label{fig:negativeRefractiveIndex}
    \end{figure}

    %\subsubsection{Anisotropic metamaterials}
    As a final example we consider negative index materials. Simultaneously negative values for $\boldsymbol{\epsilon}$ and $\boldsymbol{\mu}$ can be obtained by engineering a material at the sub-wavelength scale. Such materials are characterized by a negative refractive index~\cite{Veselago67,Pendry00}. The general algorithm presented here naturally handles negative values for both the permittivity, $\boldsymbol{\epsilon}$, and the permeability, $\boldsymbol{\mu}$. Fig.~\ref{fig:negativeRefractiveIndex} demonstrates this for a block of transparent material, yet with a negative refractive index of -1.5 ($\epsilon=-1.5$, $\mu=-1$). A Gaussian beam is incident at $30^\circ$ to the surface normal and refracts backwards into the material with an angle of $-19.5^\circ$, opposite to what is expected from Snell's law for regular glass. The wave appears to travel backwards from left to right inside the metamaterial. At the exit surface the beam couples out at $30^\circ$, again opposing Snell's law. Because there is no impedance matching in this case, some interference can be seen between the incident and reflected beam in Fig.~\ref{fig:negativeRefractiveIndex}.

  \section{Conclusion}\label{sec:conclusion}
    We have extended the computationally efficient modified Born series~\cite{Osnabrugge16,Krueger17}, to arbitrary linear electromagnetic media. This enables the accurate calculation of light propagation through birefringent, chiral, and even magnetic materials. We have implemented this numerically, and shown that a large variety of electromagnetic effects can be simulated using this method. The complete is given in Algorithm~S1, and implemented as a numerical library that can be found with working examples at \url{https://uod.box.com/v/macromax}. While there is no doubt that finite element simulations continue to be a very versatile and effective to invert the Helmholtz operator, the modified Born series has features that may make it more suitable for simulating complex samples as those found in microscopy. As the finite element method, it calculates the field due to a fixed frequency source; however, it does this through computing a series of terms that iteratively propagate the field out from the source, not entirely unlike a time-dependent calculation such as FDTD (see for example~\cite{Osnabrugge16}). Yet, unlike an FDTD calculation, the solution at a given iteration is not considered a boundary condition for the next iteration, rather an estimate of the solution to Maxwell's equations that is improved with each iteration. As such, numerical errors do not accumulate over time. Moreover, this opens the door to precalculate approximate solutions using, for instance, the beam propagation method (Supp.~Section~S1E). The modified Born series---as it was originally intended~\cite{Osnabrugge16}---is best suited for large scale problems with a limited range of susceptibilities. The modest memory requirements of the anisotropic modified Born iteration make it an ideal candidate to study light propagation in large heterogeneous tissues or materials (Supp.~Section~S1C). Our extended method allows for a wide range of material properties, including birefringence, polarization, variable permeability, chirality, and negative refractive index. This enables it to account for the birefringence and chiral effects of light traversing biological tissue. The algorithm is also expected to find use in scattering studies, where laboratory experiments are often performed on highly scattering powders that typically consist of birefringent particles such as rutile powder (TiO$_2$). The presented numerical method offers a bridge between approximate analytic predictions and experiments. In addition, such large-scale problems are not exclusive to optics and occur across wave physics, including problems involving metamaterials, to which the modified Born series can now be applied.

  \section*{Funding Information}
  	This work is supported by the Philip Leverhulme Prize from the Leverhulme Trust. SARH acknowledges financial support from the Royal Society and TATA (RPG-2016-186).

\ifthenelse{\boolean{arxiv}}{
}{
  \vspace{1em}
  	\noindent See \href{https://doi.org/\thedoi}{Supplement 1} for supporting content. %\href{http://www.opticsinfobase.org/submit/style/supplementary-materials-optica.cfm}{Supplementary Materials in Optica}.
}

  \bibliography{Common}

%  \newpage
%  % Note that this extra page will not count against page length
%  \bibliographyfullrefs{Common}